\documentclass[aip,pof,amsmath,amssymb,preprint,
superscriptaddress,
]{revtex4-1}
\ifx\pdfoutput\undefined
\usepackage{graphicx}
\else
\usepackage{graphicx}
\usepackage{epstopdf}
\fi
\usepackage[pdftitle={Yongxiang HUANG},bookmarks=true,citecolor=red,colorlinks=false]{hyperref}
\usepackage{wasysym}
\usepackage{xcolor}
\usepackage{siunitx}

\usepackage{CJK}
\newcommand{\upd}{\mathrm{\,d}}

\newcommand{\red}[1]{\textcolor{black}{#1}}
\newcommand{\blue}[1]{\textcolor{black}{#1}}

\begin{document}

\begin{CJK*}{GB}{gbsn} 
\title{Taylor dispersion in two-dimensional bacterial turbulence}

%
\author{Yongxiang Huang (»ÆÓÀÏé)}
\email{yongxianghuang@gmail.com}
\affiliation{State Key Laboratory of Marine Environmental Science \& College of Ocean and Earth Sciences,
Xiamen University, Xiamen 361102, PR China}

\author{Wenyu Ou (Å·ÎÄÓÜ)}
\affiliation{State Key Laboratory of Marine Environmental Science \& College of Ocean and Earth Sciences,
Xiamen University, Xiamen 361102, PR China}

\author{Ming Chen (³ÂÃú)}
\affiliation{State Key Laboratory of Marine Environmental Science \& College of Ocean and Earth Sciences,
Xiamen University, Xiamen 361102, PR China}

\author{Zhiming Lu (¬־Ã÷)}%
\affiliation{Shanghai Institute of Applied Mathematics and Mechanics, Shanghai Key Laboratory of Mechanics in Energy Engineering, Shanghai University,
Shanghai 200072,  China}

\author{Nan Jiang (½ªéª)}
\affiliation{Department of Mechanics, Tianjin University, 300072 Tianjin, China}

\author{Yulu Liu (ÁõÓî½)}%
\affiliation{Shanghai Institute of Applied Mathematics and Mechanics, Shanghai Key Laboratory of Mechanics in Energy Engineering, Shanghai University,
Shanghai 200072,  China}

\author{Xiang Qiu (ÇñÏè)}
\affiliation{School of Science, Shanghai Institute of Technology, Shanghai
200235,  China}

\author{Quan Zhou (ÖÜÈ«)}%
\affiliation{Shanghai Institute of Applied Mathematics and Mechanics, Shanghai Key Laboratory of Mechanics in Energy Engineering, Shanghai University,
Shanghai 200072,  China}

\date{\today}

\begin{abstract}
In this work,  single particle dispersion was analyzed for a bacterial turbulence by retrieving the virtual Lagrangian trajectory via numerical integration of the Lagrangian equation. High-order displacement functions were calculated for  cases with and without mean velocity effect. Two-regime power-law behavior for short and long time evolutions were identified experimentally, which were separated by the Lagrangian integral time.  For the case with the mean velocity effect, the experimental Hurst numbers were determined to be $0.94$ and $0.97$  for short and long times evolutions, respectively.  For the case without the mean velocity effect, the values were $0.88$ and $0.58$. Moreover, very weak intermittency correction was detected. All  measured Hurst number were above $1/2$, the value of the normal diffusion, which verifies the  super-diffusion behavior of living fluid. This behavior increases  the  efficiency of bacteria to obtain food. 

\end{abstract}

\maketitle
\end{CJK*}
\section{Introduction}

A bacterial suspension in a thin fluid can be approximated as a 2D fluid system when the considered spatial scale is larger than the fluid depth. 
In this     
active system, the energy injection and dissipation patterns are quite peculiar.  Considering  classical three-dimensional homogeneous and isotropic turbulence, the energy is  transferred from large-scale  to small-scale structures until the viscosity scale, where the kinetic energy dissipates as heat.\citep{Frisch1995}  In the bacterial turbulence,
the kinetic energy is injected into the system via a scale of the bacterial body length $R$, which is typically around a few $\mu$m, and  then transfers from small-scale to large-scale ones via an inverse cascade process.\cite{Wensink2012PNAS}  The velocity of the bacterial turbulence is also on the order of a few $\mu$m per second, and the Reynolds number is nearly zero.  Due to hydrodynamic interactions, the bacterial flow   exhibits 
 turbulent-like features, such as, long range correlation and power-law of the spectrum.\citep{Wu2000PRL,Pooley2007PRL,Ishikawa2011PRL,
Wensink2012PNAS,Dunkel2013PRL,Marchetti2013RMP,Bratanov2015PNAS,Qiu2016PRE}
For instance, \citet{Wensink2012PNAS}
observed  a dual-power-law  behavior in a quasi-2D active fluid, which was separated by the viscosity-like scale at $\ell=10R$.
\citet{Qiu2016PRE} confirmed the intermittency correction, one of the most important features of the turbulence, directly via a Hilbert-based methodology. They found a dual-power-law behavior, which is separated by a viscosity-like scale $\ell_{\nu}\simeq 10R$. The scaling behavior below the viscosity scale $\ell_{\nu}$  is more intermittent than the behavior above the scale $\ell_{\nu}$. 
This bacterial or active turbulence  is now named as ``mesoscale turbulence".\citep{Wensink2012PNAS,Bratanov2015PNAS} 

\red{Note that there is no commonly accepted unique definition of turbulent flow: it is usually identified by its main features that  a broad range of spatial and temporal scales or  many degrees of freedom are excited in the system.\citep{Groisman2000Nature} Another ``narrow definition'' of turbulence has been proposed by \citet{Gibson2004FTC} that ``turbulence is defined as an eddy-like state of fluid motion where the inertial vortex forces of the eddies are larger than any other forces that tend to dampen the eddies out."  Rotational turbulent eddies form at the viscosity scale (resp.  Kolmogorov scale), and   then they pair with neighboring ones, and these new pairs pair with neighboring pairs, and so on to generate large-scale structures.\citep{Gibson2004FTC} This is indeed the idea of the inverse energy cascade.
It seems that this definition of turbulence is applicable to the bacterial turbulence aforementioned. More discussions and examples of this ``narrow definition of turbulence'' can be found in Ref. \citep{Gibson2004FTC} and Ref.\,\footnote{See {http://journalofcosmology.com/}}.}

Single particle dispersion in disordered or turbulent flow in classical turbulence and living fluid is still fundamentally unclear. \cite{Lacasce2008PO,Bechinger2016RMP}
In  single particle dispersion,  also known as Taylor dispersion, the mean square displacement, $\mathcal{D}_2(\tau)$,  is defined as, 
\begin{equation}
\mathcal{D}_2(\tau)=\langle r_i(t,\tau)^2 \rangle_{i,t} \label{eq:MSD}
\end{equation}
where $r_i(t,\tau)=\vert \mathbf{x}_i(t,+\tau)-\mathbf{x}_i(t)\vert$ is the displacement function; $\mathbf{x}_i(t)$ is a Lagrangian trajectory; and $\tau$ is the separation time scale. According to the Taylor dispersion theory,  the following two regimes are expected,
\begin{equation}
\mathcal{D}_2(\tau)=
 \left\{
\begin{array}{lll}\label{eq:Taylor}
& \tilde{v}^2\tau^2,\quad &\textrm{when }\tau\ll T_L\\
&2\tilde{v}^2T_L\tau,\quad & \textrm{when }\tau\gg T_L
\end{array}
\right.
\end{equation}
where $T_L$ is the Lagrangian integral time scale; \red{and $\tilde{v}^2$ is the variance of Lagrangian velocity}.\citep{Taylor1921diffusion,Bouchaud1990}  The latter scaling regime is also known as normal diffusion. \cite{Metzler2000PR,Vlahos2008arXiv} Experimentally, the measured scaling exponent \red{of Eq.\,\eqref{eq:Taylor}},  $\mathcal{D}_2(\tau)\propto \tau^{\gamma_2}$,  might be different with the above mentioned values. 
For example,  \citet{Wu2000PRL}  studied   the collective dynamics of bacteria in a freely suspended soap film via the Lagrangian particle tracking technique. They found that  the measured mean displacement function of  beads demonstrates a super-diffusion (resp. $\gamma_2>1$) in short-time period and normal diffusion (resp. $\gamma_2\simeq 1$) in long-time period.   \citet{Xia2014PRL}  performed an experimental particle tracking study in a traditional two-dimensional turbulence system.  Based on the measured scaling exponent,  $\gamma_2$, a transition phenomenon of $\gamma_2$ from $2$ (\red{super-diffusion}) to $1$ (normal diffusion) was observed as  the Reynolds number increased. Based on this observation, a so-called fully developed two-dimensional turbulence was defined.  \citet{Ariel2015NC} tracked  the trajectory of individually fluorescently labeled  bacteria within such dense swarms. The authors reported a super-diffusion behavior with a measured scaling exponent $\gamma_2\simeq 1.6$, which can be further described and modeled by the L\'evy walk.  Note that for the bacterial case, the diffusion or dispersion behavior is highly dependent on many factors, such as the concentration of the bacteria, temperature and chemical gradients.

In this work, virtual particles were tracked using a numerical integration of the Lagrangian equation based on measurements of the Eulerian velocity field of bacterial turbulence via particle image velocimetry.\cite{Jullien1999PRL} The dispersion of the particles was calculated, and the results show a clear two-regime behavior, which is separated by the Lagrangian integral time. Despite the numerical value of the  scaling exponent, $\gamma_2$, the Taylor dispersion theory is evident from the obtained results. The possible intermittent correction was also checked experimentally by generalization of the mean square displacement function to  the $q$th order, e.g., $\mathcal{D}_q(\tau)=\langle r_i(t,\tau)^q \rangle_{i,t}$.  

\section{experimental data}

\begin{figure}[!htb]
\centering
\includegraphics[width=0.85\linewidth,clip]{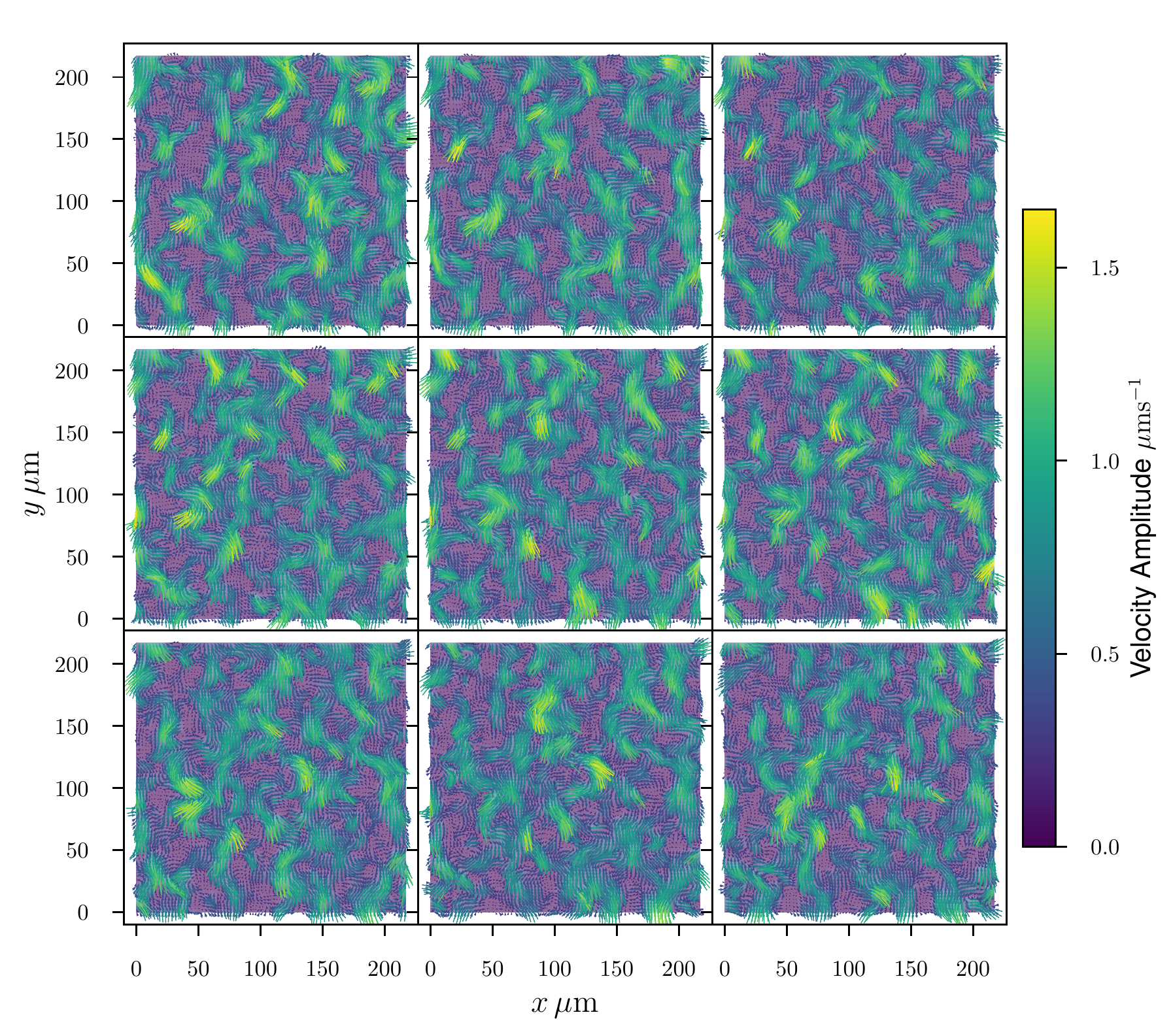}
  \caption{(Color online)  Illustration of nine successive  snapshots of instantaneous velocity vector, where the velocity amplitude is coded in color. A typical vortex structure with spatial size of $\sim 50\,\mathrm{\mu m}$ is observed.}\label{fig:snapshot}
\end{figure}

\begin{figure}[!htb]
\centering
\includegraphics[width=0.85\linewidth,clip]{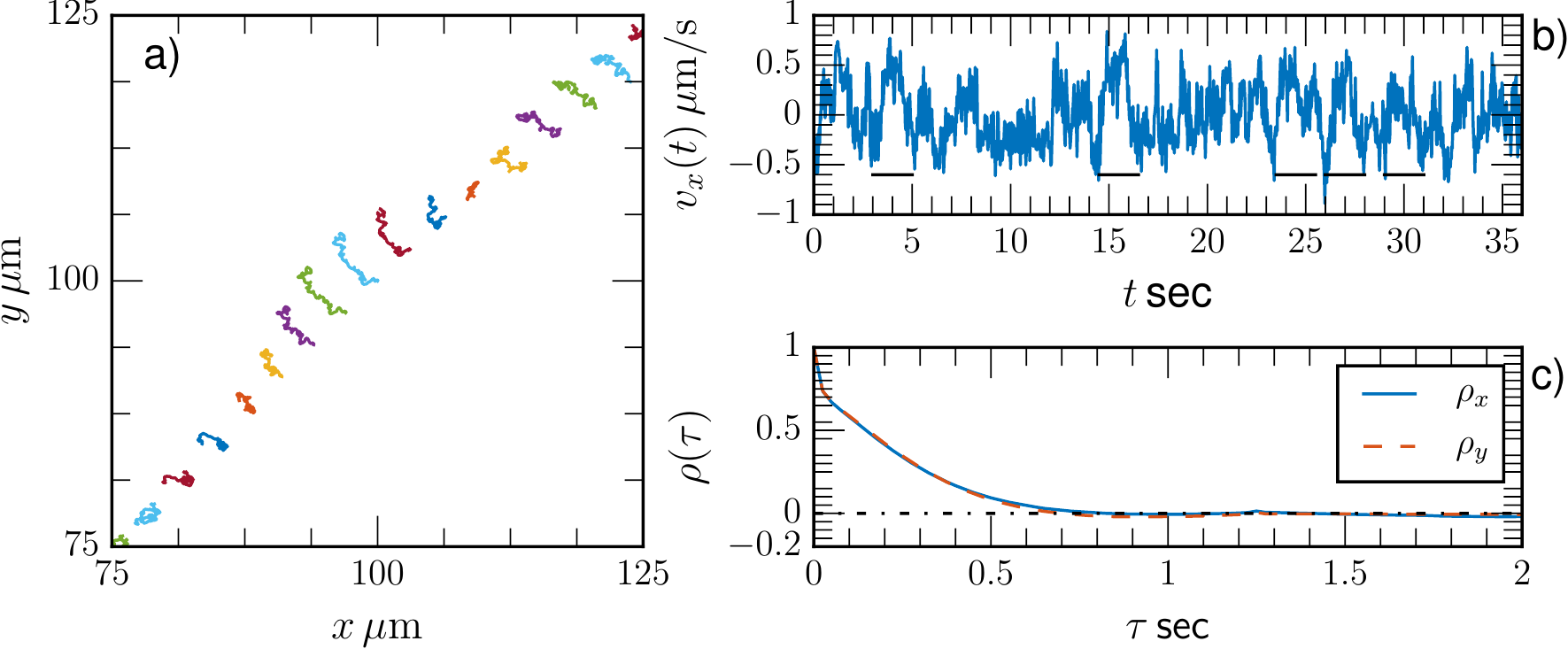}
  \caption{(Color online) a) Illustration of  Lagrangian trajectories. b) The Lagrangian velocity $v_x$ along the trajectory. \red{The horizontal solid line indicates a time span $2\,\si{sec}$.} c) Experimental autocorrelation function $\rho(\tau)$ for  Lagrangian velocities $v_x$ and $v_y$. The corresponding zero-crossing based  Lagrangian time is $\sim 0.75\,$s. }\label{fig:Trajectory}
\end{figure}

\begin{figure}[!htb]
\centering
\includegraphics[width=0.6\linewidth,clip]{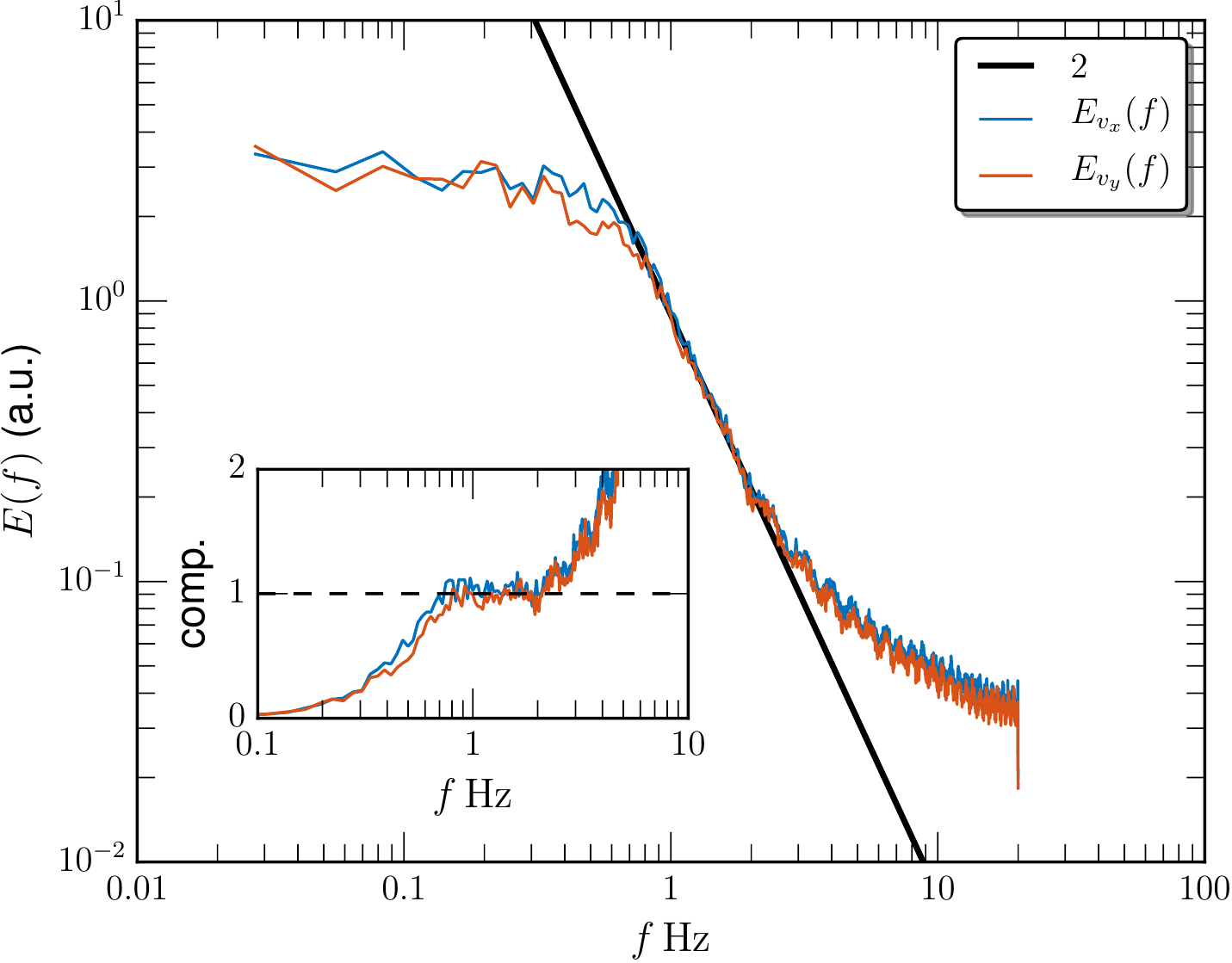}
  \caption{(Color online) Fourier power spectrum of Lagrangian velocities of ${v_x}$ and ${v_y}$. Power-law behavior is observed in the frequency range $0.6\le f\le 2\,$Hz  with a scaling exponent $2.0\pm0.06$, which is in agreement with the value predicted by the Kolmogorov-Landau theory for the conventional Lagrangian turbulence. The inset shows a compensated  curve to emphasize the observed scaling range.}\label{fig:psd}
\end{figure}

\begin{figure}[!htb]
\centering
\includegraphics[width=0.6\linewidth,clip]{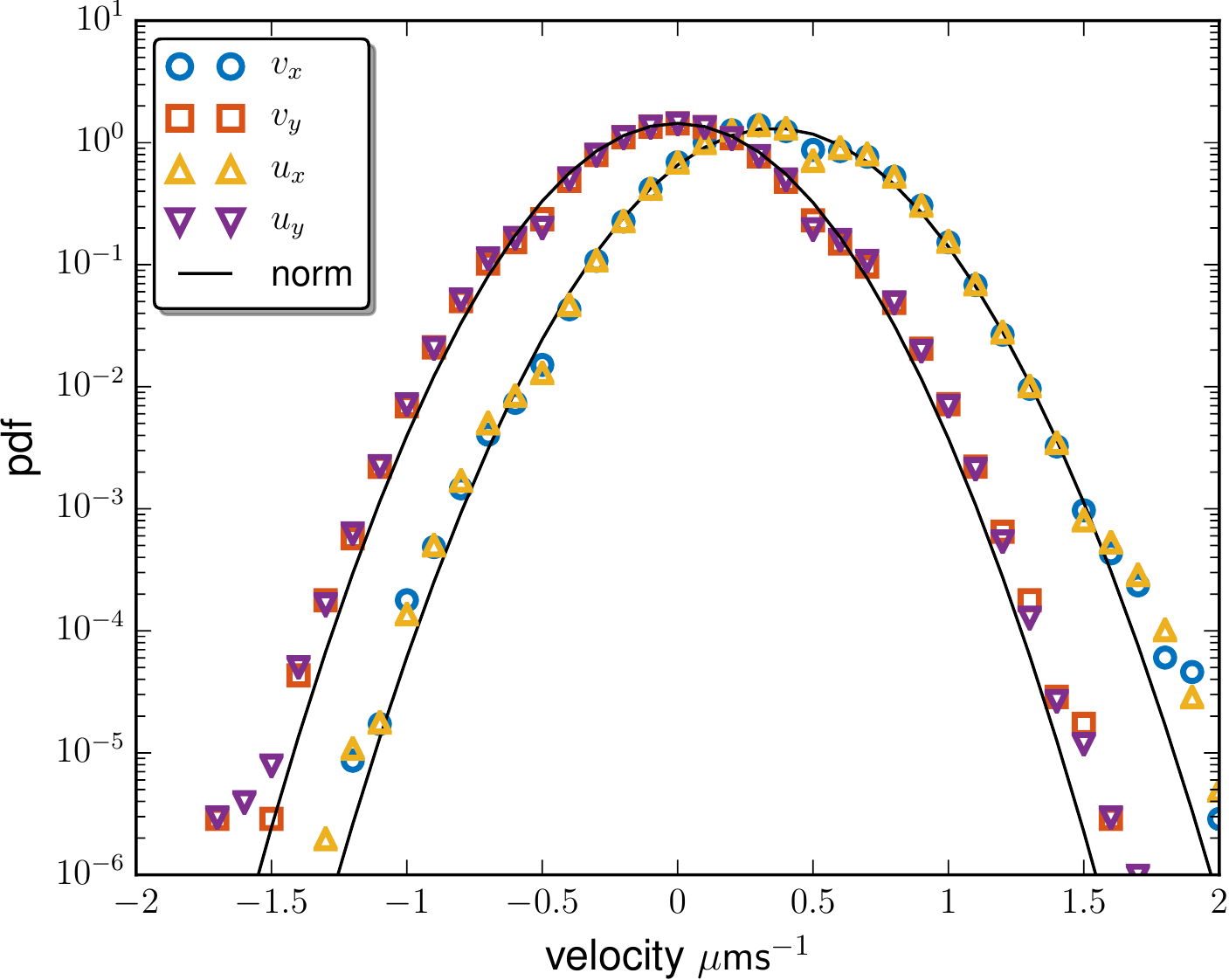}
 \caption{(Color online) Probability density function of  Eulerian ($\mathbf{u}$) and Lagrangian  velocity ($\mathbf{v}$), where normal distribution is illustrated by the solid line. }
 \label{fig:pdf}
 \end{figure}

\subsection{Experimental Setup}
The experimental data analyzed here was provided by  Professor Raymond E. Goldstein at the  Cambridge University, UK. We briefly 
recall  the main parameters of this quasi-2D experiment in a microfluidic chamber.  The bacteria used in this experiment was 
\textit{B. subtilis} with  individual  body lengths of approximately $5\,\mu$m, in which the energy is injected into the system. The 
volume filling fraction was $\phi=84\%$ with particle number $N=9968$ and aspect ratio $a=5$ (the ratio between the bacterial 
body length $R$ and the body diameter). The quasi-2D microfluidic chamber had  a vertical height, $H_c$, less than or equal to the 
individual body length of \textit{B. subtilis} (approximately 5$\,\mu$m). With these  parameters, the flow entered a turbulent 
phase.\cite{Wensink2012PNAS} The  particle image velocimetry (PIV) had a measurement area of  
$217\mathrm{\mu m}\times 217\mathrm{\mu m}$.  The image resolution was  $700\,$pix$\times700\,$pix with a conversion rate of 
$0.31\,\mathrm{\mu m}/$pix and a frame rate of  
$40\,$Hz. The commercial PIV software Dantec Flow Manager was used to extract the flow field component with a moving window size 
of $32\,$pix$\times 32\,$pix and $75\%$ overlap. This resulted in an  $84\times 84$ velocity vector and a total of 1441 
snapshots, corresponding to a period of $\sim36\,$seconds. Therefore, a spatial structure  size $\ell$ larger than $H_c$ can be 
treated as a two-dimensional system.  The mean and root-mean-square (rms.) velocities were determined to be $
\overline{\mathbf{u}}=(0.36,0)\,\mathrm{\mu ms^{-1}}$ and $\mathbf{u}^{'}=(0.30,0.29)\,\mathrm{\mu ms^{-1}}$, respectively, and the 
corresponding turbulent intensity was around $\mathbf{u}'/\overline{\mathbf{u}}_x\simeq 82\%$.  The PIV uncertainty  was  
less than $1\%$ for the second-order statistics. More details of this database can be found in Ref.\,\onlinecite{Wensink2012PNAS}.

Figure \ref{fig:snapshot} shows nine successive snapshots of the instantaneous velocity vector, where the velocity amplitude is coded in color.  A typical flow structure with a spatial size around $\sim 50\,\mathrm{\mu m}$ is visually evident and  has been recognized as a fluid-viscosity-like scale.\cite{Qiu2016PRE} The flow is smooth in space due to fluctuations in the high wave number (for structures smaller than $50\,\mathrm{\mu m}$) that will be quickly damped by the viscosity.

\subsection{Numerical Tracking Algorithm}
Using the measured Eulerian velocity field, $\mathbf{u}(\mathbf{x},t)$, we numerically integrated the Lagrangian equation as follows,
\begin{equation}
\frac{\upd \mathbf{x}}{\upd t} =\mathbf{v}(\mathbf{x},t)
\end{equation}
where $\mathbf{v}(\mathbf{x},t)$ is the Lagrangian velocity; and $\mathbf{v}(\mathbf{x},t)$ is equal to  the Eulerian velocity at the same position, e.g., $\mathbf{v}(\mathbf{x},t)=\mathbf{u}(\mathbf{x},t)$. 
A second-order Adams-Bashforth method was employed in the time scheme, while a two-dimensional spline interpolation scheme was used to retrieve the Lagrangian velocity not on the grid point.  The virtual particles were assumed to be free with the bacteria body, or in other words, they could freely penetrate the bacteria body. 
Initially, $2500$ virtual fluid particles were seeded with a uniform distribution in the range $(x,y)\in [0,217]\,\mathrm{\mu m}$. If a particle moved beyond the experimental area, time integration  stopped.  For one realization, about $\sim$100 virtual particles exceeded the boundary, and ten realizations were performed.    In totally, we obtained $2400\times 1441\times10 \simeq 34,584,000$  (number of virtual particles per realization $\times$ number of  snapshots $\times$ number of realizations) velocity vectors to ensure a good statistics.

Figure \ref{fig:Trajectory}\,a) illustrates  several trajectories that show a complex behavior. Figure \ref{fig:Trajectory}\,b) shows the Lagrangian velocity $v_x(t)$ along the Lagrangian trajectory, where a large structure fluctuation is visible  around $\sim 2\,\si{sec}$.  From there, we estimated the Lagrangian velocity correlation function as, 
\begin{equation}
\rho(\tau)=\frac{\langle \tilde{v}(t+\tau)\tilde{v}(t) \rangle_{t} }{\sigma^2}
\end{equation} 
where $\tilde{v}(t)=v(t)-\langle v(t)\rangle_t$ is a centered Lagrangian velocity; $\langle \,\rangle_t$ is time average; and $\sigma$ is the root-mean-square Lagrangian velocity. Figure \ref{fig:Trajectory}\,c) displays the experimental $\rho(\tau)$ for $v_x(t)$ and $v_y(t)$, which show long range correlation  in time.  The Lagrangian       time scale, $T_L$, is defined as, 
\begin{equation}
T_L=\int_0^{\infty}\rho(\tau)\upd \tau
\end{equation}
Due to the finite time measurement, it is difficult to apply the above definition directly. Therefore, we used the first zero-crossing time as the Lagrangian integral time, i.e., $T_L=\tau_0$, where $\rho(\tau_0)=0$. The experimental value of $T_L$ is found to be $\sim0.75\,\mathrm{sec}$;  thus we  expect two dispersion regimes to be separated by the Lagrangian time scale, $T_L$.

Figure \ref{fig:psd} shows the measured Fourier power spectrum for the Lagrangian velocities $v_x$ and $v_y$.  Power-law behavior,  $E(f)\sim f^{-\beta}$, was observed in the range $0.6\le f\le2\,\mathrm{Hz}$, corresponding to the  time scale range $0.5\le t\le1.7\,\mathrm{sec}$. A compensated curve using the measured scaling exponent is also shown in the inset in Fig.\,\ref{fig:psd} to emphasize the power-law behavior, where a clear plateau confirms the existence of the power-law behavior. The measured scaling exponent is $\beta=2.0\pm0.06$, \red{where the error is provided by the $95\%$ fitting confidence level of the least square fit algorithm. This scaling value}  is coincidently in a good agreement with the value predicted by the Kolmogorov-Landau theory for  hydrodynamical Lagrangian turbulence.\cite{Falkovich2012PoF,Huang2013PRE} Note that, due to the existence of vortex trapping events in the conventional three-dimensional Lagrangian turbulence, this value is difficult to obtain even turbulent flows with high Reynolds numbers.\cite{Falkovich2012PoF}

Figure \ref{fig:pdf} shows the measured probability density function (pdf) of Eulerian velocity,  $\mathbf{u}$, and Lagrangian velocity, $\mathbf{v}$, where the normal distribution is represented by a solid line. As seen in the figure, the measured pdfs of  Eulerian and Lagrangian velocities agree well with the normal distribution.\cite{Wensink2012PNAS}

\section{Results}
 \begin{figure}[!htb]
\centering
\includegraphics[width=1\linewidth,clip]{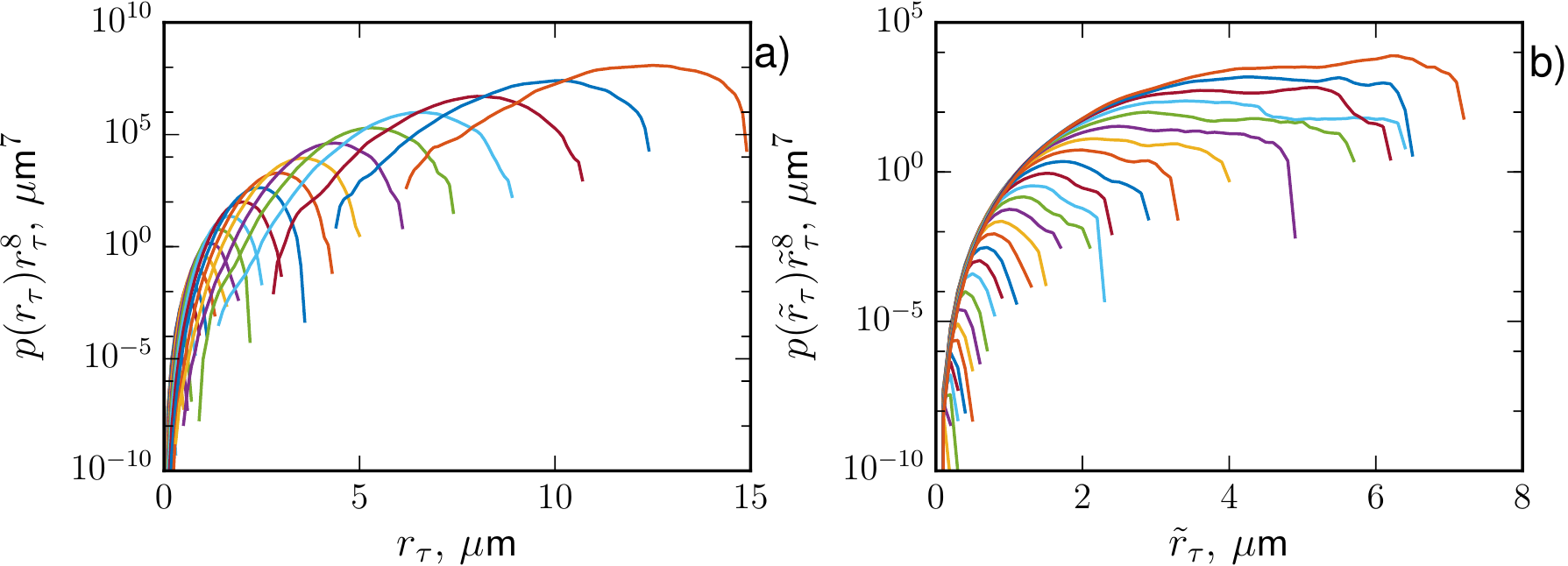}
 \caption{(Color online)  Convergence test of $8$th-order displacement function $\mathcal{D}_8(\tau)=\int p(r_{\tau})r_{\tau}^8\upd r_{\tau}$, where $p(r_{\tau})$ is the experimental pdf of  the displacement function with the separation time scale $\tau$.
 The corresponding integral kernel: a) with mean velocity effect, and b) without mean velocity effect.  It confirms the statistical convergence for all separation scales at least up to the statistical order $q=8$.}
 \label{fig:Cpdf}
 \end{figure}
 
\begin{figure}[!htb]
\centering
\includegraphics[width=1\linewidth,clip]{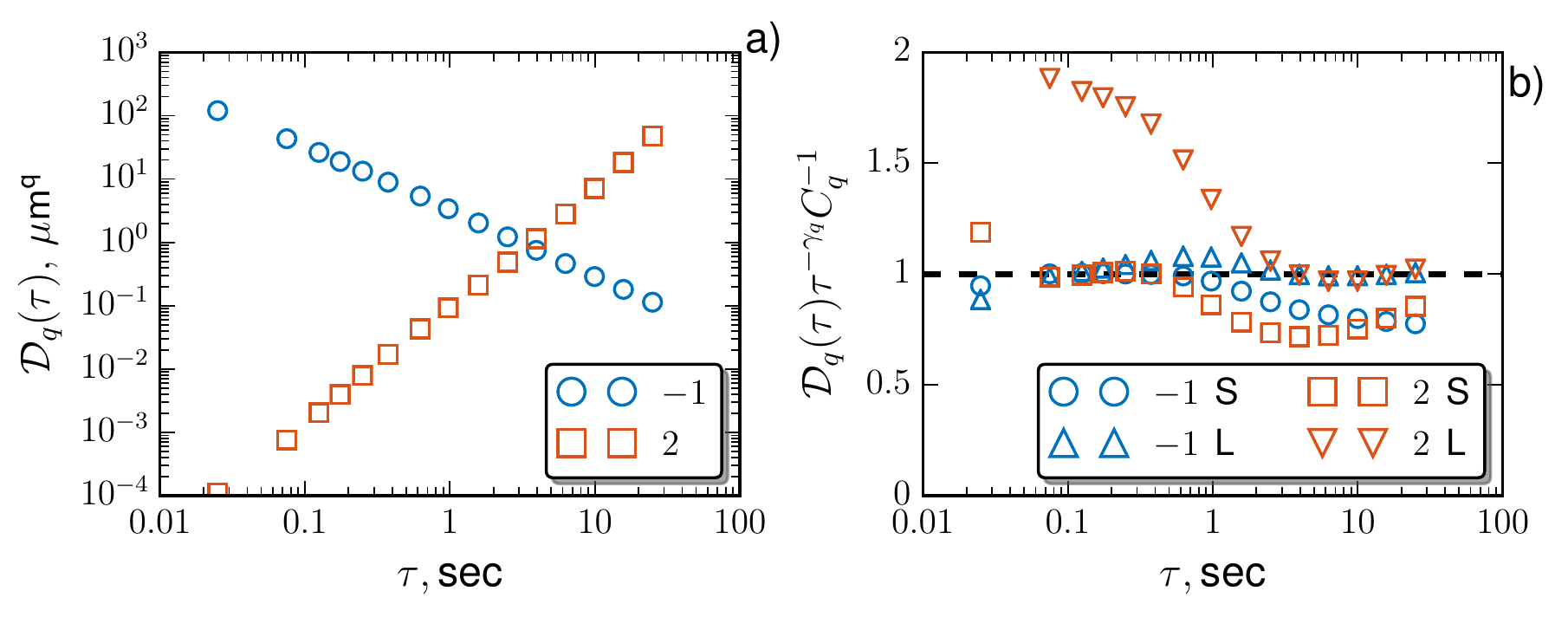}
 \caption{(Color online) a) Measured $q$th-order  displacement function $\mathcal{D}_q(\tau)$ for statistical order  $q=-1$ ($\ocircle$) and $2$ ($\square$). b) The compensated curve to emphasize the power-law behavior in the range $0.05\le \tau\le 0.5\,$sec and $2\le \tau \le 31\,$sec using measured scaling exponents $\gamma_2^S=1.87\pm 0.01$ and $\gamma^L_2=1.93\pm0.01$, respectively, and the corresponding fitting constant, $C_q$.}
 \label{fig:QMD}
 \end{figure}

 \begin{figure}[!htb]
\centering
\includegraphics[width=1\linewidth,clip]{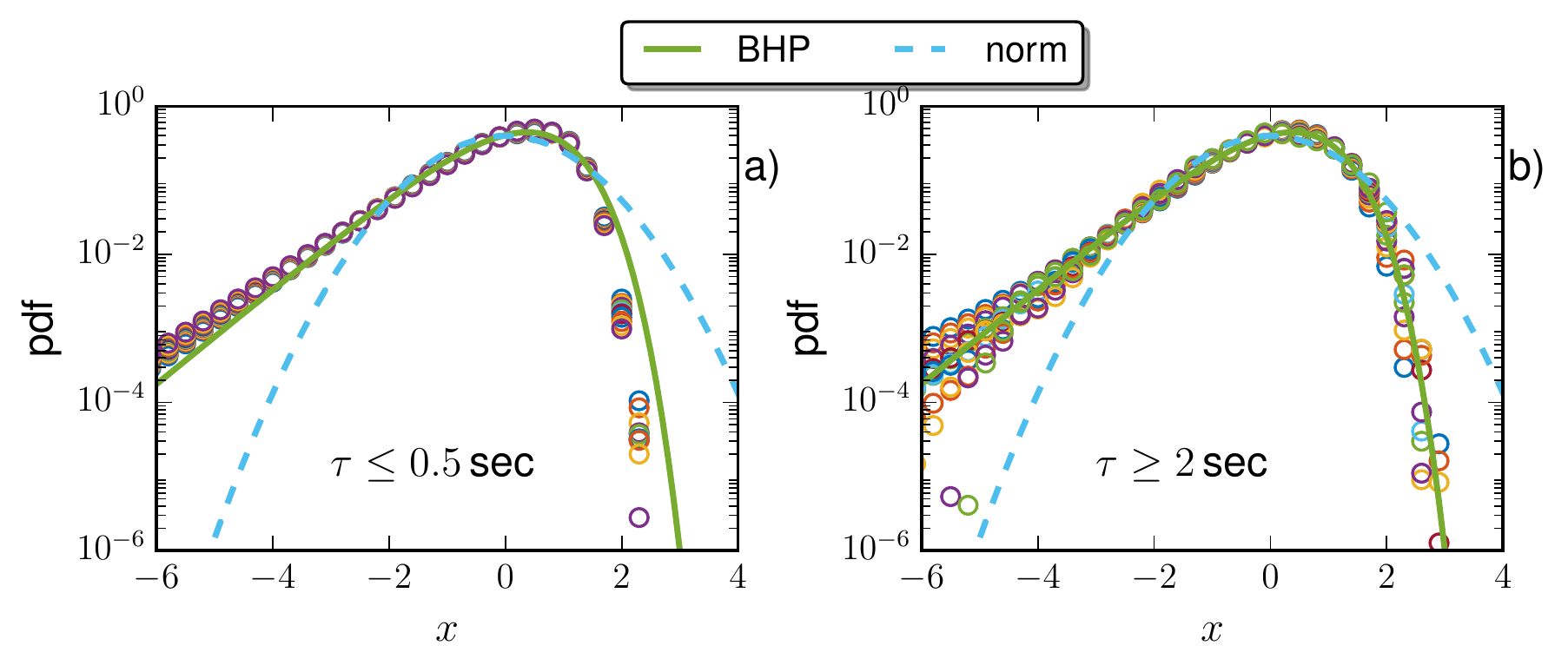}
 \caption{(Color online) Experimental pdf of the centered $x$, where $x=(x_{\tau}-\langle x_{\tau} \rangle)/\sigma$ and $x_{\tau}=\log_{10}(r(\tau))$ for a) $\tau\le 0.5\,$sec and b) $\tau\ge 2\,$sec. For comparison, the Bramwell-Holdsworth-Pinton formula (thick solid line) and the normal distribution (dashed line) are also shown.    }
 \label{fig:Rpdf}
 \end{figure}
 
 \begin{figure}[!htb]
\centering
\includegraphics[width=1\linewidth,clip]{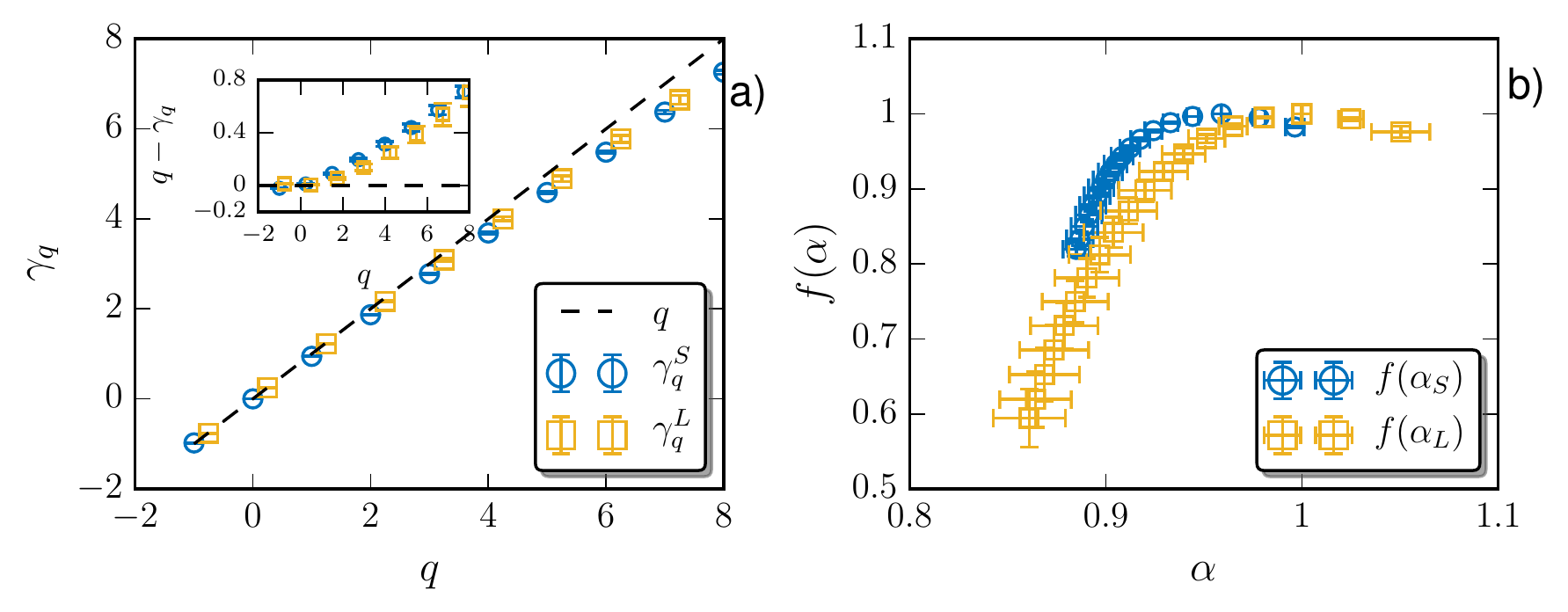}
 \caption{(Color online) a) Measured scaling exponents, $\gamma_q$, for short and long time evolutions. For comparison, the scaling $\gamma_q=q$ is represented  by a dashed line. The inset shows the deviation from the linear relation $\gamma_q=q$. b) The corresponding singularity spectrum, $f(\alpha)$, versus $\alpha$.  \red{The error is provided by the $95\%$ fitting confidence level.}}
 \label{fig:QScaling}
 \end{figure}

\subsection{Convergence test}
To analyze the convergence of statistics in this work, we first considered the $q$th-order displacement function, which is defined as,
\begin{equation}
\mathcal{D}_q(\tau)=\langle r_i(t+\tau)^q\rangle_{i,t}
\end{equation}
which can be re-written as, 
\begin{equation}
\mathcal{D}_q(\tau)=\int_0^{\infty} p(r_{\tau})r_{\tau}^q\upd r_{\tau}
\end{equation}
where $p(r_{\tau})$ is the experimental pdf  for the separation time scale, $\tau$; and $p(r_{\tau})r_{\tau}^q$ is the so-called $q$th-order integral-kernel.   Figure \ref{fig:Cpdf} shows the measured integral-kernels for a) with and b) without the mean velocity effect by removing the persistent velocity effect from the displacement function, i.e., $\tilde{r}_i(\tau)=\vert \mathbf{x}_i(t+\tau)-\mathbf{x}_i(t)-V\tau \vert $, where $V=\langle v(t)\rangle_{t}$ is the mean Lagrangian velocity. 
  Visually, the measured curve initially increased  and then  curved downward, which  indicates a good convergence of the statistics at least up to the statistical order $q=8$. We then considered the statistics in the range $-1\le q \le 8$ and  time scale in the range $0.05\le \tau\le 31\,$sec.

\subsection{Taylor dispersion with mean velocity effect}
Figure \ref{fig:QMD}\,a) shows the experimental $q$th-order displacement function for $q=-1$ ($\ocircle$) and $q=2$ ($\square$). 
The power-law behavior is evident, as expected, and  can be further identified by two regimes, e.g., $0.05\le \tau\le 0.5\,$sec, and $2\le\tau\le 31\,$sec, which are separated by the above estimated Lagrangian time scale, $T_L=0.75\,\mathrm{sec}$.
Figure \ref{fig:QMD}\,b) displays the corresponding compensated curve using the fitted scaling exponent, $\gamma_2$, to emphasize the observed power-law behavior, in which a clear plateau confirms the existence of the power-law behavior.  The measured scaling exponents  for the case $q=2$  were determined to be $\gamma^S_{2}=1.87\pm0.01$ and $ \gamma_{2}^L=1.93\pm0.01$, \red{where the error is provided by the $95\%$ fitting confidence level}. The former value is close to the value $\gamma_2=2$, which was predicted by the Taylor dispersion theory for a short time evolution. Due to the existence of the mean  velocity effect, the latter scaling exponent is also close to $2$.

Experimental pdfs for the normalized displacement functions, $x=(x_{\tau}-\langle x_{\tau}\rangle)/\sigma$ and  $x_{\tau}=\log_{10}(r(\tau))$, were calculated with a bin width of $0.1$. 
Figure \ref{fig:Rpdf} shows the measured pdfs   for a) $\tau\le 0.5\,$sec and b) $\tau\ge 2\,$sec.  Except for slight scattering at the right ($x\ge 1$) and left ($x\le -4$) tails, 
they collapsed with each other on their own regimes, indicating the scale invariance property.   For comparison, the normal distribution and the Bramwell-Holdsworth-Pinton (BHP) formula,\citep{Bramwell1998Nature} are  displayed as dashed and solid lines, respectively. 
The BHP formula is written as
\begin{equation}
\Pi(y)=K(e^{y-e^y})^a,\,y=b(x-s),a=\pi/2,
\end{equation}
where parameters $b=0.938$, and $K=2.14$ were obtained numerically from a previous study. \citep{Bramwell2000PRL}
This formula  was first introduced to characterize rare fluctuations in turbulence and critical phenomena.  The measured pdfs for small-scale separation time  ($\tau\le 0.5\,$sec) agrees with the BHP formula on the range $-4\le x\le0.6$. For large-scale separation time ($\tau\le 2\,$sec), the pdfs agree well  with the BHP when $x\ge -4$.

We then estimated the scaling exponents on these two scaling regimes for $-1\le q\le 8$. Figure \ref{fig:QScaling}\,a) shows the measured scaling exponents, $\gamma_q$, for both short  ($\ocircle$) and long time ($\square$) evolutions, where $\gamma_q=q$ predicted by the Taylor dispersion theory for short time evolutions  is  shown as a dashed line. 
The error bar indicates a $95\%$ fitting confidence interval. 
Experimentally, the measured scaling exponents, $\gamma^S_q$ and $\gamma^L_q$, are close to each other since  the mean Lagrangian  velocity effect is preserved in the numerical integral algorithm, but deviate from the prediction of the Taylor dispersion theory (inset of Fig.\,\ref{fig:QScaling}\,a). A nonlinear $q$-dependence  of these curves implies a potential intermittency correction. 
 To characterize the potential intermittency effect, we introduce here 
a lognormal formula to fit the measured scaling exponents, which is written as, 
\begin{equation}
\gamma(q)=qH-\frac{\mu}{2}\left(q^2H^2-qH\right)
\end{equation}
where $H$ is the Hurst number and $\mu$ is the intermittency parameter. \citep{Li2014PhysicaA}  This lognormal formula is a generalization of the classical one proposed by Kolmogorov in 1960s for the hydrodynamic turbulence with $H=1/3$.  For a given $H$, a larger value of $\mu$ is, a more intermittenter process. The measured parameters $H_S=0.94\pm0.01$ and $\mu_S=0.01\pm0.001$, and $H_L=0.97\pm0.01$ and $\mu_L=0.02\pm0.001$ show a very weak intermittent correction, \red{where the error is provided by the $95\%$ fitting confidence level}. 

Note that for the lognormal formula,  both $H$ and $\mu$ are taken as free parameters.  The measured intermittency parameter, $\mu$, depends on  $H$. To overcome this difficulty,  we introduce here a singularity spectrum, which is defined via a Legendre transform,
\begin{equation}
f(\alpha)=\min_q\{q\alpha-\gamma_q+1\},\quad \alpha(q)=\frac{\upd \gamma_q}{\upd q}
\end{equation}
where $\alpha$ is the  multifractal intensity; and $f(\alpha)$ is the singularity spectrum. Experimentally, a wider $\alpha$ and $f(\alpha)$ refer to  a more intermittent field.  Figure \ref{fig:QScaling}\,b) shows the measured $f(\alpha)$ versus $\alpha$, where  a weak intermittent correction exists for both small and large time scales. Moreover, the singularity spectrum for large-scale time separations seems to be more intermittent than that for small-scale one. 

\subsection{Taylor dispersion without mean velocity effect}

\begin{figure}
\centering
\includegraphics[width=0.85\linewidth,clip]{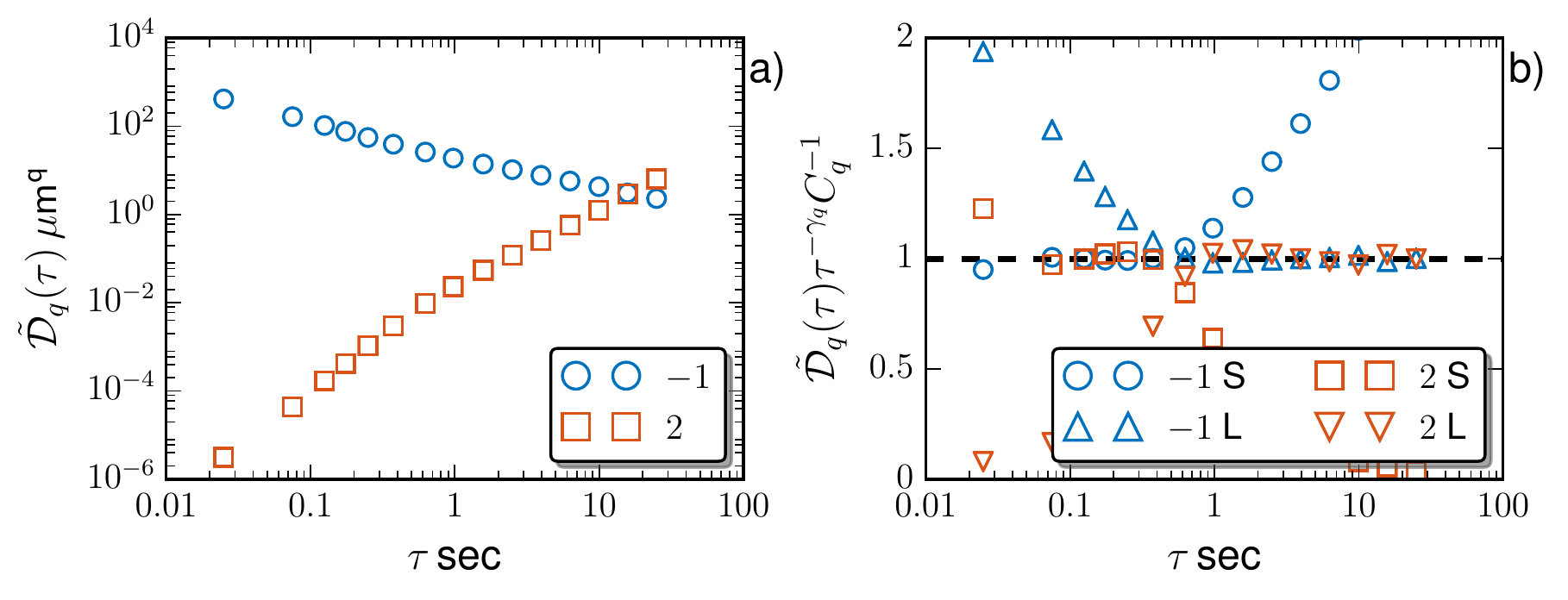}
\caption{(Color online) a) Measured $q$th-order mean displacement function, $\tilde{\mathcal{D}}_q(\tau)$, for statistical order $q=-1$ ($\ocircle$) and $2$ ($\square$) without mean velocity effect. b) The compensated curve to emphasize the power-law behavior in the ranges $0.05\le \tau\le 0.5\,$sec and $2\le \tau \le 31\,$sec. }
\label{fig:QMD2}
\end{figure}

\begin{figure}
\centering
\includegraphics[width=0.85\linewidth,clip]{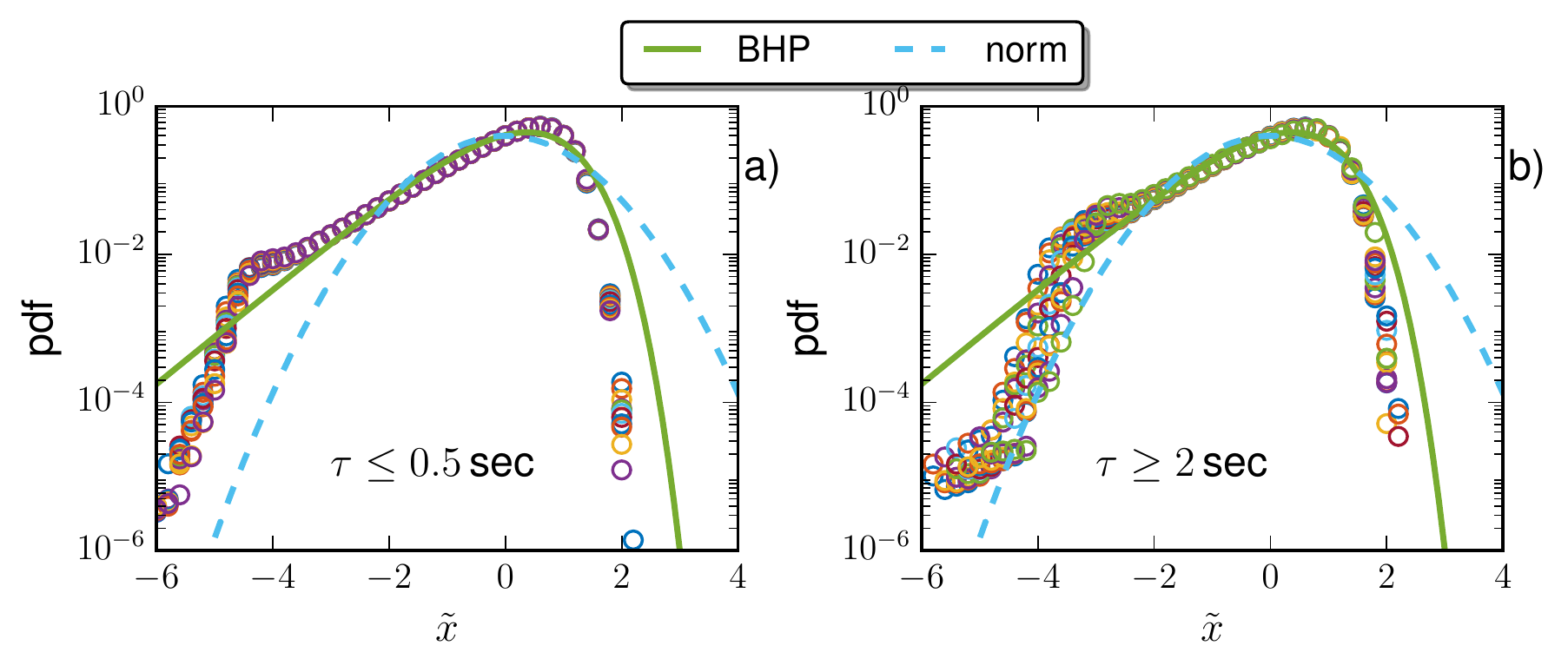}
\caption{(Color online) Experimental normalized pdfs of the centered displacement $\tilde{x}$, where $\tilde{x}=(\tilde{x}_{\tau}-\langle \tilde{x}_{\tau} \rangle)/\sigma$ and $\tilde{x}_{\tau}=\log_{10}(\tilde{r}(\tau))$ for a) $\tau\le 0.5\,$sec and b) $\tau\ge 2\,$sec. For comparison, the BHP formula and the normal distribution are also shown. }
\label{fig:Rpdf2}
\end{figure}
 \begin{figure}
\centering
\includegraphics[width=1\linewidth,clip]{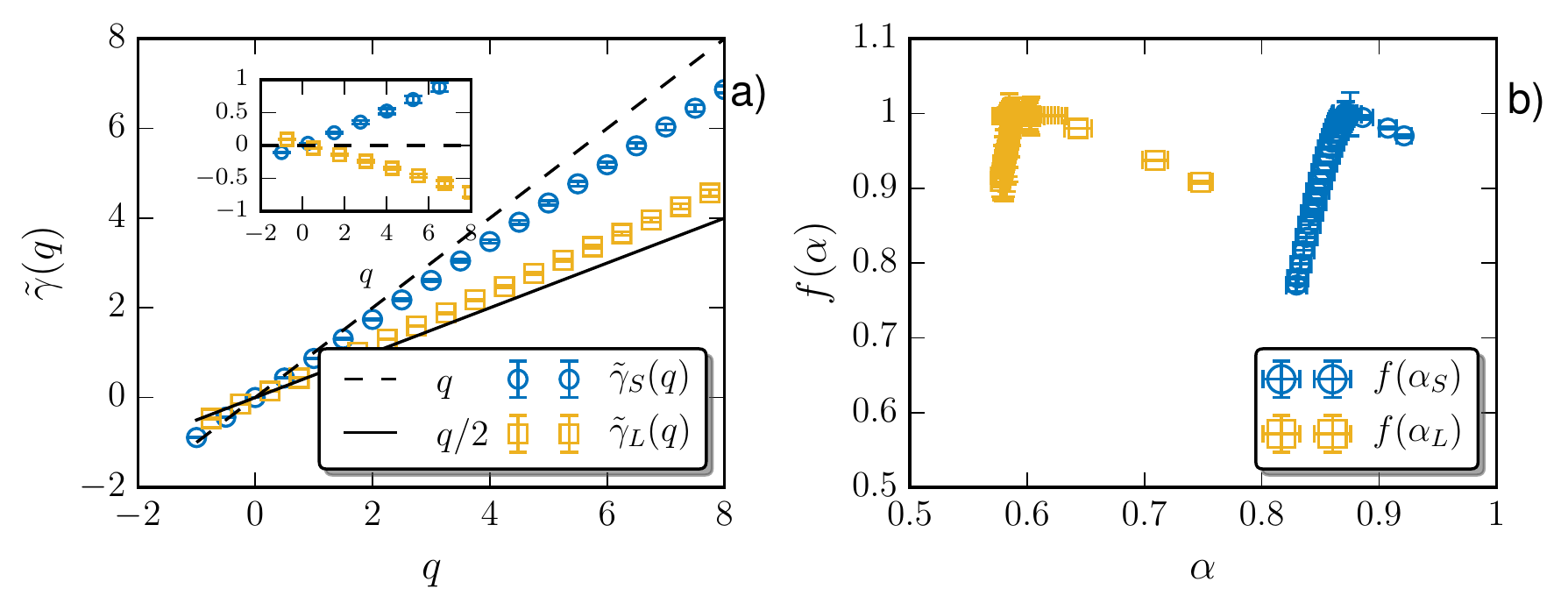}
 \caption{(Color online) a) Measured scaling exponents $\tilde{\gamma}_q$ for short and long time evolutions without mean velocity effect. For comparison, the scaling $\tilde{\gamma}_q=q$ and $\tilde{\gamma}_q=q/2$ are illustrated as dashed and solid lines. b) The corresponding singularity spectrum, $f(\alpha)$, versus $\alpha$.}
 \label{fig:QScaling2}
 \end{figure}

To exclude the  mean velocity effect, a displacement function   that omits this effect is defined as,
\begin{equation}
\mathcal{\tilde{D}}_q(\tau)=\langle \tilde{r}_i(t+\tau)^q \rangle_{i,t} 
\end{equation}
where $\tilde{r}_i(t+\tau)=\vert \mathbf{x}_i(t+\tau)-\mathbf{x}_i(t) -V\tau\vert$; and $V=\langle v(t)\rangle_{t}$ is the mean Lagrangian velocity. This formula represents a dispersion observed on a moving frame with a constant velocity, $V$. Figure \ref{fig:QMD2}\,a) shows the measured $\tilde{\mathcal{D}}_q(\tau)$ for statistical order $q=-1$ ($\ocircle$) and $q=2$ ($\square$). Power-law behavior is observed with two different regimes in the time  ranges $0.05\le \tau\le 0.5\,$sec and $2\le \tau\le 31\,$sec with scaling exponents $\tilde{\gamma}^S_2=1.74\pm0.02$ and $\tilde{\gamma}^L_2=1.16\pm0.01$, \red{where the error is provided by the $95\%$ fitting confidence level}.  The former scaling exponent is smaller than the one predicted by the Taylor dispersion theory for short time evolutions, and is similar  to the one reported by \citet{Ariel2015NC}. The latter scaling exponent for long time evolutions  is slight larger than the value of $1$, indicating a super-diffusion. This result reveals the effects of   bacterial turbulence.  The compensated curve is shown in Fig.\,\ref{fig:QMD2}\,b) to emphasize the power-law behavior. A clear plateau confirms the existence of the power-law. 

Figure \ref{fig:Rpdf2} shows the experimental pdfs $p(\tilde{x})$ of $\tilde{x}=(\tilde{x}_{\tau}-\langle \tilde{x}_{\tau} \rangle)/\sigma$ and $\tilde{x}_{\tau}=\log_{10}(\tilde{r}(\tau))$ for a) $0.05\le \tau\le 0.5\,$sec and b) $2\le \tau \le 31\,$sec. Visually, these differ greatly  from  the pdfs obtained with the BHP formula or normal distribution. Moreover, we discovered that the pdfs for small (resp. $0.05\le \tau \le 0.5\,$sec) and large (resp. $2\le \tau\le31\,$sec) times collapse with each other when $-4\le x$.  The collapse implies that, if the intermittency correction exists, their strengths could be the same for these two different regimes. 

Figure \ref{fig:QScaling2}\,a) shows the measured scaling exponents, $\tilde{\gamma}_q$, for both small ($\ocircle$) and large ($\square$) time evolutions.  For comparison, values predicted by Taylor dispersion theory  are also shown. Visually, the scaling exponent $\tilde{\gamma}_q^S$ in short time evolutions is below the theoretical one when $q\ge 0$, while the expnoent in long time evolution is above the theoretical one when $q\ge 0$, indicating  \blue{super-diffusion}.  
 The corresponding deviation from the theoretical values are shown in the inset in Fig.\ref{fig:QScaling2}\,a). Note that these curves are almost linear against $q$, indicating that the possible intermittency correction is possibly negligible.
The lognormal formula fitting
provides $\tilde{H}_S=0.88\pm0.01$ and $\tilde{\mu}_S=0.01\pm0.001$, and
 $\tilde{H}_L=0.58\pm0.01$ and $\tilde{\mu}_L=-0.01\pm0.005$, \red{where the error is provided by the $95\%$ fitting confidence level}. These values suggest a very weak intermittency correction, confirming the observations from Fig.\,\ref{fig:QScaling2}\,a). Figure \ref{fig:QScaling2}\,b) shows the measured singularity spectrum, $f(\alpha)$, versus $\alpha$ where
the intermittency correction could be ignored for both short and long time evolutions.

\section{Discussion}

\begin{figure}
\centering
\includegraphics[width=0.6\linewidth,clip]{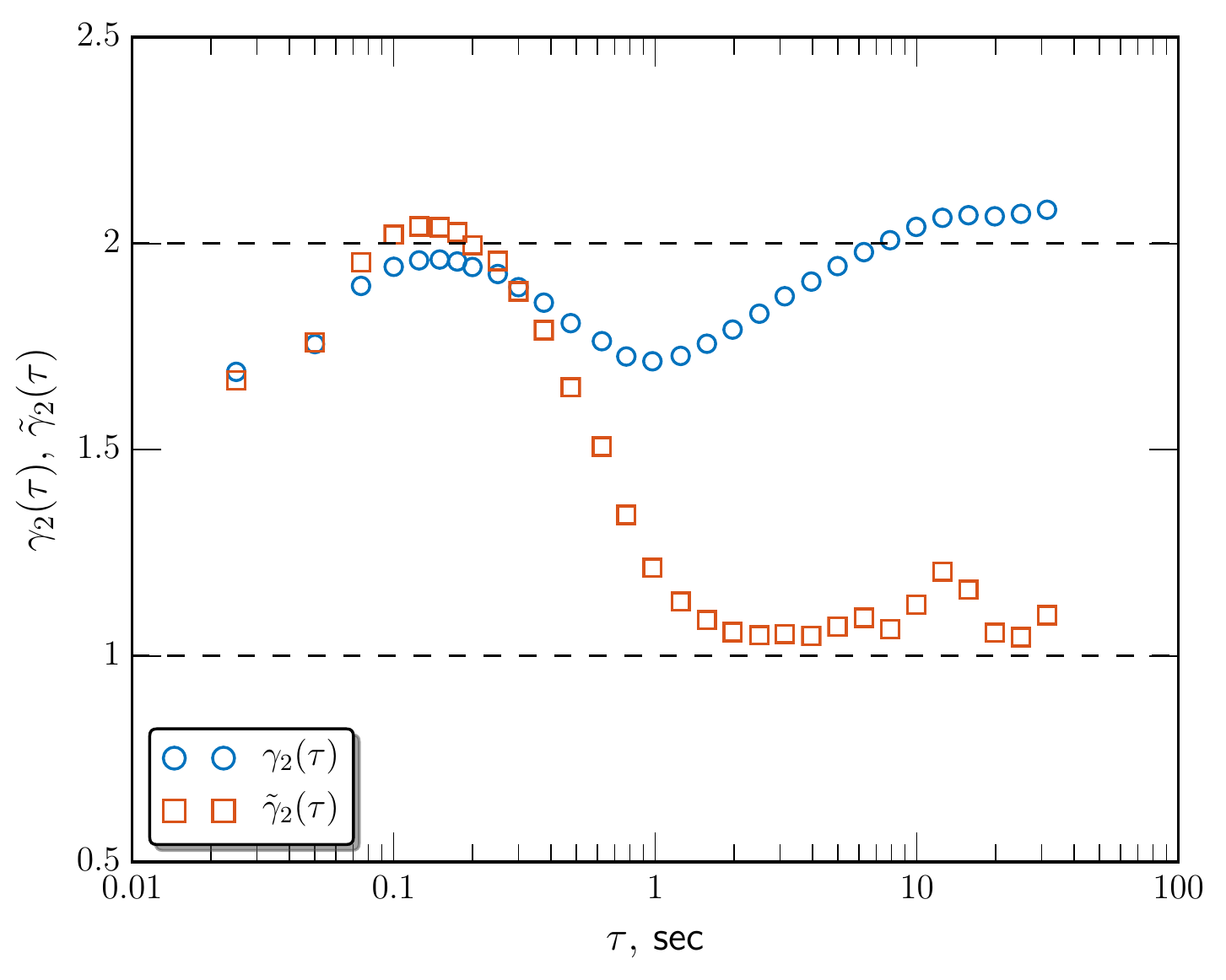}
\caption{(Color online) Experimental local scaling exponent $\gamma_2(\tau)$ (resp. $\tilde{\gamma}_2(\tau)$) for the mean square displacement for $\mathcal{D}_2(\tau)$  (resp. $\mathcal{\tilde{D}}_2(\tau)$). The lack of a  plateau implies the absence of a pure power-law behavior.}
\label{fig:localscaling}
\end{figure}

\begin{figure}
\centering
\includegraphics[width=0.85\linewidth,clip]{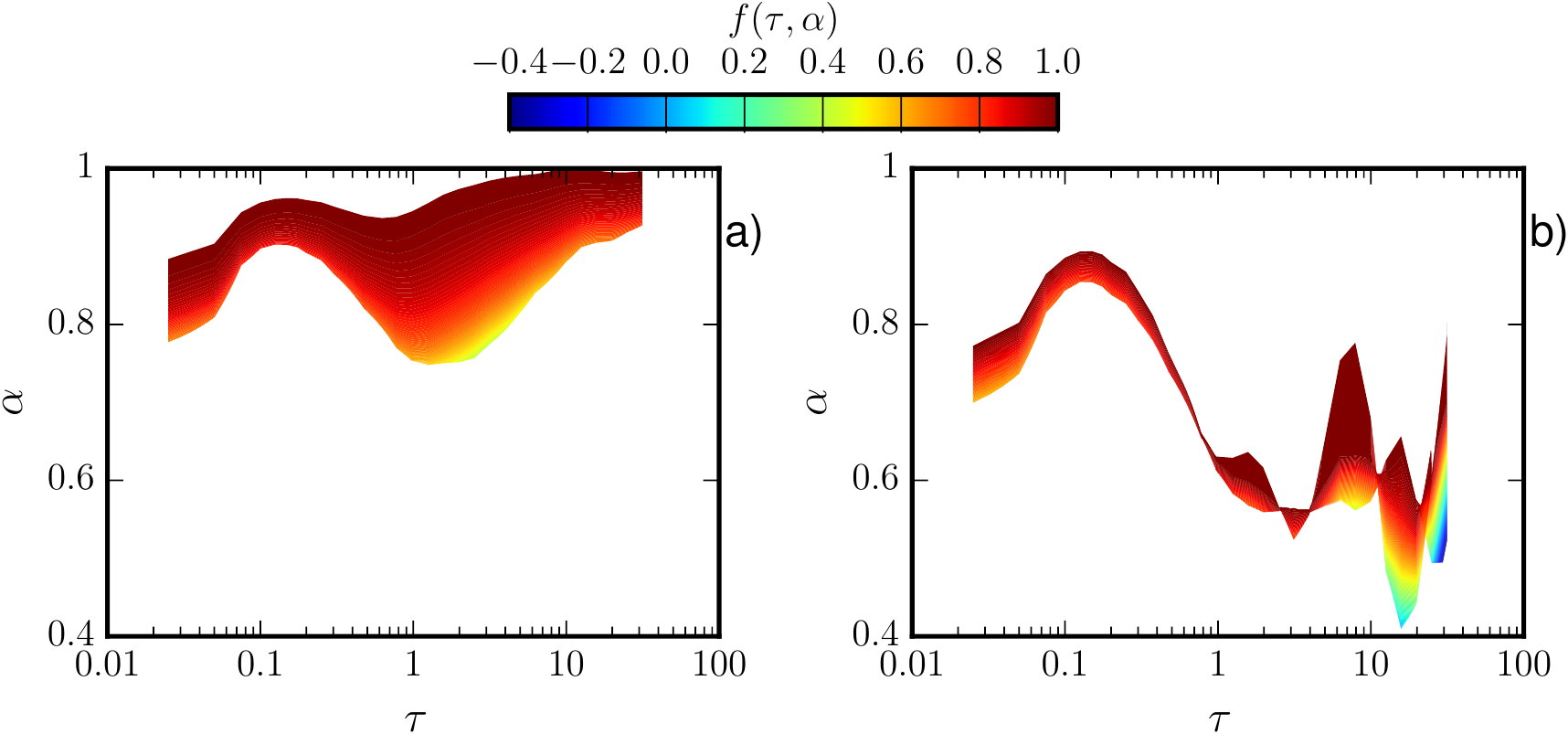}
\caption{(Color online)  Experimental local singularity spectrum,  $f(\tau,\alpha)$, a) with  and b) without mean velocity effect.  For comparison convenience, measured $f(\tau,\alpha)$ is  coded by the same color map.}
\label{fig:localsingulairty}
\end{figure}

The measured scaling exponents, $\gamma_q$, may vary for different fitting ranges. This is because that a pure power-law behavior is difficult to retrieve, and this can be seen from a local scaling exponent, $\gamma_2(\tau)$, which is defined as, 
\begin{equation}
\gamma_2(\tau)=\frac{\upd \log_{10} \mathcal{D}_2(\tau)}{\upd \log_{10}\tau}
\end{equation}
This exponent can be generalized for a $q$th-order displacement function $\mathcal{D}_q(\tau)$. 
Figure \ref{fig:localscaling} shows the measured $\gamma_2(\tau)$ (resp. $\tilde{\gamma}_2(\tau)$), where the lack of a clear plateau implies the absence of pure power-law behavior. In other words, the experimental scaling exponent, $\gamma_2$, might depend on the choice of the fitting range. 
To avoid this vagueness, we introduce here a local singularity spectrum, 
\begin{equation}
f(\tau,\alpha)=\min_q\left\{q\alpha(\tau,q)-\gamma_q(\tau)+1 \right\},\,\alpha(\tau,q)=\frac{\upd \gamma_q(\tau)}{\upd q}
\end{equation}
where $\alpha(\tau,q)$ is the local multifractal intensity; and $f(\tau,\alpha)$ is the local singularity spectrum.
Figure \ref{fig:localsingulairty} shows a contour plot of the measured local singularity spectrum $f(\tau,\alpha)$ respectively for a) with and b) without the mean velocity effect. For visual convenience, the plots are coded in the same color map. 
 Visually, two different regimes are visible with either a small variation  of $\alpha(\tau,q)$ or $f(\tau,\alpha)$, indicating that  the intermittency correction  is very weak if it exists.  This is consistent  with the direct measurement of the intermittency parameter, $\mu\simeq 0.01$.
 
 The measured Hurst number, $H$, either with or without  mean velocity effect are different than  the ones predicted by the Taylor dispersion theory. This difference may be an effect of the active movement of bacteria, which are aided by super-diffusion to capture food.\citep{Ariel2015NC} 
 Although the two regimes are experimentally evident, this result is a consequence of the  turbulent motion of such bacterial turbulent flow.

Finally, we would like to provide some comments on the statistical uncertainty of this study. As mentioned above, the statistical uncertainty of the PIV measurement is less than $1\%$ for the second-order moment.\citep{Wensink2012PNAS} \red{We have tested  for different  algorithms  for both time advance  and  spatial  interpolation, and found that the individual trajectories starting from the same initial positions vary slightly.} However, the statistical moments of displacement functions are the same, which is partially due to a rather smooth fluid field. Further,  the displacement function, $\mathcal{D}_q(\tau)$, is indeed a structure-function of the Lagrangian trajectory. Therefore, $V_{\tau}=\mathcal{D}_1(\tau)/\tau$ can be treated as a coarse-grained mean velocity, which is less influenced by the intermittency effect that has been found for the Eulerian velocity field.\citep{Qiu2016PRE} 
\red{We also note that some velocity vectors from 10 realizations are not statistically independent. But it does not change the conclusion of this work.}
However, a true particle tracking experiment should be done with carefully designed  as a means to directly check the dispersion relations.

 \section{Conclusion}
In summary,  single particle dispersion was analyzed for a bacterial turbulence by  numerical integration of the Lagrangian equation. 
A second-order
 Adams-Bashforth scheme in time and a two-dimensional
spline interpolation scheme in space were used, and  $q$th-order displacement functions  with and without the mean velocity effect were calculated. 
The results show a two-regime behavior respectively in  short and long time evolutions with  corresponding regimes in the range $0.05\le \tau\le 0.5\,$sec and $2\le \tau \le 31\,$sec, respectively.  With the mean velocity effect, the experimental  pdfs of the displacement function was fitted via the BHP formula, and the measured scaling exponents, $\gamma_q$, was found to  deviate from the one  predicted by the Taylor dispersion theory, which   can be understood as an effect  of the active dynamic system. 
Moreover, the corresponding Hurst numbers were determined to be $H_S=0.94$ and $H_L=0.97$. When excluding the  mean velocity effect, the measured pdfs of displacement function collapse with each other and were  different from the BHP formula, while the measured scaling exponents, $\gamma_q$, deviated from the theoretical  prediction. More precisely, in short time evolutions ($0.05\le \tau\le 0.5\,$sec), the measured Hurst number was  $\tilde{H}_S=0.88$,  smaller than the value of the theoretical prediction.  In long time evolutions  ($2\le \tau\le 31\,$sec), the measured Hurst number was $\tilde{H}_L=0.58$ and is larger than the value predicted by the Taylor dispersion theory. All these measured Hurst numbers indicate a super-diffusion behavior, which is relevant for bacteria to gain food more efficiently.  Furthermore, the experimental results show a weak  intermittency effect.
  Excluding the results obtained for the Hurst numbers, the two-regime behavior predicted by the Taylor dispersion theory is verified for this active turbulence system.

\begin{acknowledgments}

This work is partially sponsored by the National Natural Science Foundation of China under Grant (No.  11332006, 11572203), and  the Fundamental Research Funds for the Central Universities (Grant No. 20720150069 (Y.H.)).  Y.H. is also supported  by the Sino-French (NSFC-CNRS) joint project  (No. 11611130099, NSFC China, and PRC 2016-2018 LATUMAR ``Turbulence lagrangienne: \'etudes num\'eriques et applications environnementales marines",  CNRS, France).  
 We thank professor  Raymond E. Goldstein  for  providing us the experiment data, which can be found at \,\footnote{See {http://damtp.cam.ac.uk/user/gold/datarequests.html}}.
A \textsc{Matlab} source code  package to realize the numerical tracking algorithm is available at \,\footnote{See {https://github.com/lanlankai}}.
\end{acknowledgments}


\begin{thebibliography}{28}%
\makeatletter
\providecommand \@ifxundefined [1]{%
 \@ifx{#1\undefined}
}%
\providecommand \@ifnum [1]{%
 \ifnum #1\expandafter \@firstoftwo
 \else \expandafter \@secondoftwo
 \fi
}%
\providecommand \@ifx [1]{%
 \ifx #1\expandafter \@firstoftwo
 \else \expandafter \@secondoftwo
 \fi
}%
\providecommand \natexlab [1]{#1}%
\providecommand \enquote  [1]{``#1''}%
\providecommand \bibnamefont  [1]{#1}%
\providecommand \bibfnamefont [1]{#1}%
\providecommand \citenamefont [1]{#1}%
\providecommand \href@noop [0]{\@secondoftwo}%
\providecommand \href [0]{\begingroup \@sanitize@url \@href}%
\providecommand \@href[1]{\@@startlink{#1}\@@href}%
\providecommand \@@href[1]{\endgroup#1\@@endlink}%
\providecommand \@sanitize@url [0]{\catcode `\\12\catcode `\$12\catcode
  `\&12\catcode `\#12\catcode `\^12\catcode `\_12\catcode `\%12\relax}%
\providecommand \@@startlink[1]{}%
\providecommand \@@endlink[0]{}%
\providecommand \url  [0]{\begingroup\@sanitize@url \@url }%
\providecommand \@url [1]{\endgroup\@href {#1}{\urlprefix }}%
\providecommand \urlprefix  [0]{URL }%
\providecommand \Eprint [0]{\href }%
\providecommand \doibase [0]{http://dx.doi.org/}%
\providecommand \selectlanguage [0]{\@gobble}%
\providecommand \bibinfo  [0]{\@secondoftwo}%
\providecommand \bibfield  [0]{\@secondoftwo}%
\providecommand \translation [1]{[#1]}%
\providecommand \BibitemOpen [0]{}%
\providecommand \bibitemStop [0]{}%
\providecommand \bibitemNoStop [0]{.\EOS\space}%
\providecommand \EOS [0]{\spacefactor3000\relax}%
\providecommand \BibitemShut  [1]{\csname bibitem#1\endcsname}%
\let\auto@bib@innerbib\@empty
\bibitem [{\citenamefont {Frisch}(1995)}]{Frisch1995}%
  \BibitemOpen
  \bibfield  {author} {\bibinfo {author} {\bibfnamefont {U.}~\bibnamefont
  {Frisch}},\ }\href@noop {} {\emph {\bibinfo {title} {{Turbulence: the legacy
  of AN Kolmogorov}}}}\ (\bibinfo  {publisher} {Cambridge University Press},\
  \bibinfo {year} {1995})\BibitemShut {NoStop}%
\bibitem [{\citenamefont {Wensink}\ \emph {et~al.}(2012)\citenamefont
  {Wensink}, \citenamefont {Dunkel}, \citenamefont {Heidenreich}, \citenamefont
  {Drescher}, \citenamefont {Goldstein}, \citenamefont {L{\"o}wen},\ and\
  \citenamefont {Yeomans}}]{Wensink2012PNAS}%
  \BibitemOpen
  \bibfield  {author} {\bibinfo {author} {\bibfnamefont {H.~H.}\ \bibnamefont
  {Wensink}}, \bibinfo {author} {\bibfnamefont {J.}~\bibnamefont {Dunkel}},
  \bibinfo {author} {\bibfnamefont {S.}~\bibnamefont {Heidenreich}}, \bibinfo
  {author} {\bibfnamefont {K.}~\bibnamefont {Drescher}}, \bibinfo {author}
  {\bibfnamefont {R.~E.}\ \bibnamefont {Goldstein}}, \bibinfo {author}
  {\bibfnamefont {H.}~\bibnamefont {L{\"o}wen}}, \ and\ \bibinfo {author}
  {\bibfnamefont {J.~M.}\ \bibnamefont {Yeomans}},\ }\bibfield  {title}
  {\enquote {\bibinfo {title} {Meso-scale turbulence in living fluids},}\
  }\href@noop {} {\bibfield  {journal} {\bibinfo  {journal} {PNAS}\ }\textbf
  {\bibinfo {volume} {109}},\ \bibinfo {pages} {14308--14313} (\bibinfo {year}
  {2012})}\BibitemShut {NoStop}%
\bibitem [{\citenamefont {Wu}\ and\ \citenamefont
  {Libchaber}(2000)}]{Wu2000PRL}%
  \BibitemOpen
  \bibfield  {author} {\bibinfo {author} {\bibfnamefont {X.-L.}\ \bibnamefont
  {Wu}}\ and\ \bibinfo {author} {\bibfnamefont {A.}~\bibnamefont {Libchaber}},\
  }\bibfield  {title} {\enquote {\bibinfo {title} {Particle diffusion in a
  quasi-two-dimensional bacterial bath},}\ }\href@noop {} {\bibfield  {journal}
  {\bibinfo  {journal} {Phys. Rev. Lett.}\ }\textbf {\bibinfo {volume} {84}},\
  \bibinfo {pages} {3017} (\bibinfo {year} {2000})}\BibitemShut {NoStop}%
\bibitem [{\citenamefont {Pooley}, \citenamefont {Alexander},\ and\
  \citenamefont {Yeomans}(2007)}]{Pooley2007PRL}%
  \BibitemOpen
  \bibfield  {author} {\bibinfo {author} {\bibfnamefont {C.~M.}\ \bibnamefont
  {Pooley}}, \bibinfo {author} {\bibfnamefont {G.~P.}\ \bibnamefont
  {Alexander}}, \ and\ \bibinfo {author} {\bibfnamefont {J.~M.}\ \bibnamefont
  {Yeomans}},\ }\bibfield  {title} {\enquote {\bibinfo {title} {{Hydrodynamic
  interaction between two swimmers at low Reynolds number}},}\ }\href@noop {}
  {\bibfield  {journal} {\bibinfo  {journal} {Phys. Rev. Lett.}\ }\textbf
  {\bibinfo {volume} {99}},\ \bibinfo {pages} {228103} (\bibinfo {year}
  {2007})}\BibitemShut {NoStop}%
\bibitem [{\citenamefont {Ishikawa}\ \emph {et~al.}(2011)\citenamefont
  {Ishikawa}, \citenamefont {Yoshida}, \citenamefont {Ueno}, \citenamefont
  {Wiedeman}, \citenamefont {Imai},\ and\ \citenamefont
  {Yamaguchi}}]{Ishikawa2011PRL}%
  \BibitemOpen
  \bibfield  {author} {\bibinfo {author} {\bibfnamefont {T.}~\bibnamefont
  {Ishikawa}}, \bibinfo {author} {\bibfnamefont {N.}~\bibnamefont {Yoshida}},
  \bibinfo {author} {\bibfnamefont {H.}~\bibnamefont {Ueno}}, \bibinfo {author}
  {\bibfnamefont {M.}~\bibnamefont {Wiedeman}}, \bibinfo {author}
  {\bibfnamefont {Y.}~\bibnamefont {Imai}}, \ and\ \bibinfo {author}
  {\bibfnamefont {T.}~\bibnamefont {Yamaguchi}},\ }\bibfield  {title} {\enquote
  {\bibinfo {title} {Energy transport in a concentrated suspension of
  bacteria},}\ }\href@noop {} {\bibfield  {journal} {\bibinfo  {journal} {Phys.
  Rev. Lett.}\ }\textbf {\bibinfo {volume} {107}},\ \bibinfo {pages} {028102}
  (\bibinfo {year} {2011})}\BibitemShut {NoStop}%
\bibitem [{\citenamefont {Dunkel}\ \emph {et~al.}(2013)\citenamefont {Dunkel},
  \citenamefont {Heidenreich}, \citenamefont {Drescher}, \citenamefont
  {Wensink}, \citenamefont {B{\"a}r},\ and\ \citenamefont
  {Goldstein}}]{Dunkel2013PRL}%
  \BibitemOpen
  \bibfield  {author} {\bibinfo {author} {\bibfnamefont {J.}~\bibnamefont
  {Dunkel}}, \bibinfo {author} {\bibfnamefont {S.}~\bibnamefont {Heidenreich}},
  \bibinfo {author} {\bibfnamefont {K.}~\bibnamefont {Drescher}}, \bibinfo
  {author} {\bibfnamefont {H.~H.}\ \bibnamefont {Wensink}}, \bibinfo {author}
  {\bibfnamefont {M.}~\bibnamefont {B{\"a}r}}, \ and\ \bibinfo {author}
  {\bibfnamefont {R.~E.}\ \bibnamefont {Goldstein}},\ }\bibfield  {title}
  {\enquote {\bibinfo {title} {Fluid dynamics of bacterial turbulence},}\
  }\href@noop {} {\bibfield  {journal} {\bibinfo  {journal} {Phys. Rev. Lett.}\
  }\textbf {\bibinfo {volume} {110}},\ \bibinfo {pages} {228102} (\bibinfo
  {year} {2013})}\BibitemShut {NoStop}%
\bibitem [{\citenamefont {Marchetti}\ \emph {et~al.}(2013)\citenamefont
  {Marchetti}, \citenamefont {Joanny}, \citenamefont {Ramaswamy}, \citenamefont
  {Liverpool}, \citenamefont {Prost}, \citenamefont {Rao},\ and\ \citenamefont
  {Simha}}]{Marchetti2013RMP}%
  \BibitemOpen
  \bibfield  {author} {\bibinfo {author} {\bibfnamefont {M.~C.}\ \bibnamefont
  {Marchetti}}, \bibinfo {author} {\bibfnamefont {J.~F.}\ \bibnamefont
  {Joanny}}, \bibinfo {author} {\bibfnamefont {S.}~\bibnamefont {Ramaswamy}},
  \bibinfo {author} {\bibfnamefont {T.~B.}\ \bibnamefont {Liverpool}}, \bibinfo
  {author} {\bibfnamefont {J.}~\bibnamefont {Prost}}, \bibinfo {author}
  {\bibfnamefont {M.}~\bibnamefont {Rao}}, \ and\ \bibinfo {author}
  {\bibfnamefont {R.~A.}\ \bibnamefont {Simha}},\ }\bibfield  {title} {\enquote
  {\bibinfo {title} {Hydrodynamics of soft active matter},}\ }\href@noop {}
  {\bibfield  {journal} {\bibinfo  {journal} {Rev. Mod. Phys.}\ }\textbf
  {\bibinfo {volume} {85}},\ \bibinfo {pages} {1143} (\bibinfo {year}
  {2013})}\BibitemShut {NoStop}%
\bibitem [{\citenamefont {Bratanov}, \citenamefont {Jenko},\ and\ \citenamefont
  {Frey}(2015)}]{Bratanov2015PNAS}%
  \BibitemOpen
  \bibfield  {author} {\bibinfo {author} {\bibfnamefont {V.}~\bibnamefont
  {Bratanov}}, \bibinfo {author} {\bibfnamefont {F.}~\bibnamefont {Jenko}}, \
  and\ \bibinfo {author} {\bibfnamefont {E.}~\bibnamefont {Frey}},\ }\bibfield
  {title} {\enquote {\bibinfo {title} {New class of turbulence in active
  fluids},}\ }\href@noop {} {\bibfield  {journal} {\bibinfo  {journal} {PNAS}\
  }\textbf {\bibinfo {volume} {112}},\ \bibinfo {pages} {15048--15053}
  (\bibinfo {year} {2015})}\BibitemShut {NoStop}%
\bibitem [{\citenamefont {Qiu}\ \emph {et~al.}(2016)\citenamefont {Qiu},
  \citenamefont {Ding}, \citenamefont {Huang}, \citenamefont {Chen},
  \citenamefont {Lu}, \citenamefont {Liu},\ and\ \citenamefont
  {Zhou}}]{Qiu2016PRE}%
  \BibitemOpen
  \bibfield  {author} {\bibinfo {author} {\bibfnamefont {X.}~\bibnamefont
  {Qiu}}, \bibinfo {author} {\bibfnamefont {L.}~\bibnamefont {Ding}}, \bibinfo
  {author} {\bibfnamefont {Y.}~\bibnamefont {Huang}}, \bibinfo {author}
  {\bibfnamefont {M.}~\bibnamefont {Chen}}, \bibinfo {author} {\bibfnamefont
  {Z.}~\bibnamefont {Lu}}, \bibinfo {author} {\bibfnamefont {Y.}~\bibnamefont
  {Liu}}, \ and\ \bibinfo {author} {\bibfnamefont {Q.}~\bibnamefont {Zhou}},\
  }\bibfield  {title} {\enquote {\bibinfo {title} {Intermittency measurement in
  two-dimensional bacterial turbulence},}\ }\href@noop {} {\bibfield  {journal}
  {\bibinfo  {journal} {Phys. Rev. E}\ }\textbf {\bibinfo {volume} {93}},\
  \bibinfo {pages} {062226} (\bibinfo {year} {2016})}\BibitemShut {NoStop}%
\bibitem [{\citenamefont {Groisman}\ and\ \citenamefont
  {Steinberg}(2000)}]{Groisman2000Nature}%
  \BibitemOpen
  \bibfield  {author} {\bibinfo {author} {\bibfnamefont {A.}~\bibnamefont
  {Groisman}}\ and\ \bibinfo {author} {\bibfnamefont {V.}~\bibnamefont
  {Steinberg}},\ }\bibfield  {title} {\enquote {\bibinfo {title} {Elastic
  turbulence in a polymer solution flow},}\ }\href@noop {} {\bibfield
  {journal} {\bibinfo  {journal} {Nature}\ }\textbf {\bibinfo {volume} {405}},\
  \bibinfo {pages} {53--55} (\bibinfo {year} {2000})}\BibitemShut {NoStop}%
\bibitem [{\citenamefont {Gibson}(2004)}]{Gibson2004FTC}%
  \BibitemOpen
  \bibfield  {author} {\bibinfo {author} {\bibfnamefont {C.~H.}\ \bibnamefont
  {Gibson}},\ }\bibfield  {title} {\enquote {\bibinfo {title} {The first
  turbulence and first fossil turbulence},}\ }\href@noop {} {\bibfield
  {journal} {\bibinfo  {journal} {Flow Turb. Comb.}\ }\textbf {\bibinfo
  {volume} {72}},\ \bibinfo {pages} {161--179} (\bibinfo {year}
  {2004})}\BibitemShut {NoStop}%
\bibitem [{Note1()}]{Note1}%
  \BibitemOpen
  \bibinfo {note} {See {http://journalofcosmology.com/}}\BibitemShut {NoStop}%
\bibitem [{\citenamefont {LaCasce}(2008)}]{Lacasce2008PO}%
  \BibitemOpen
  \bibfield  {author} {\bibinfo {author} {\bibfnamefont {J.}~\bibnamefont
  {LaCasce}},\ }\bibfield  {title} {\enquote {\bibinfo {title} {{Statistics
  from Lagrangian observations}},}\ }\href@noop {} {\bibfield  {journal}
  {\bibinfo  {journal} {Prog. Oceanograph.}\ }\textbf {\bibinfo {volume}
  {77}},\ \bibinfo {pages} {1--29} (\bibinfo {year} {2008})}\BibitemShut
  {NoStop}%
\bibitem [{\citenamefont {Bechinger}\ \emph {et~al.}(2016)\citenamefont
  {Bechinger}, \citenamefont {Di~Leonardo}, \citenamefont {L{\"o}wen},
  \citenamefont {Reichhardt}, \citenamefont {Volpe},\ and\ \citenamefont
  {Volpe}}]{Bechinger2016RMP}%
  \BibitemOpen
  \bibfield  {author} {\bibinfo {author} {\bibfnamefont {C.}~\bibnamefont
  {Bechinger}}, \bibinfo {author} {\bibfnamefont {R.}~\bibnamefont
  {Di~Leonardo}}, \bibinfo {author} {\bibfnamefont {H.}~\bibnamefont
  {L{\"o}wen}}, \bibinfo {author} {\bibfnamefont {C.}~\bibnamefont
  {Reichhardt}}, \bibinfo {author} {\bibfnamefont {G.}~\bibnamefont {Volpe}}, \
  and\ \bibinfo {author} {\bibfnamefont {G.}~\bibnamefont {Volpe}},\ }\bibfield
   {title} {\enquote {\bibinfo {title} {Active {Brownian} particles in complex
  and crowded environments},}\ }\href@noop {} {\bibfield  {journal} {\bibinfo
  {journal} {Rev. Mod. Phys.}\ }\textbf {\bibinfo {volume} {88}},\ \bibinfo
  {pages} {045006} (\bibinfo {year} {2016})}\BibitemShut {NoStop}%
\bibitem [{\citenamefont {Taylor}(1921)}]{Taylor1921diffusion}%
  \BibitemOpen
  \bibfield  {author} {\bibinfo {author} {\bibfnamefont {G.~I.}\ \bibnamefont
  {Taylor}},\ }\bibfield  {title} {\enquote {\bibinfo {title} {Diffusion by
  continuous movements},}\ }\href@noop {} {\bibfield  {journal} {\bibinfo
  {journal} {Proc. R. Soc. Lond.}\ }\textbf {\bibinfo {volume} {20}},\ \bibinfo
  {pages} {196--211} (\bibinfo {year} {1921})}\BibitemShut {NoStop}%
\bibitem [{\citenamefont {Bouchaud}\ and\ \citenamefont
  {Georges}(1990)}]{Bouchaud1990}%
  \BibitemOpen
  \bibfield  {author} {\bibinfo {author} {\bibfnamefont {J.-P.}\ \bibnamefont
  {Bouchaud}}\ and\ \bibinfo {author} {\bibfnamefont {A.}~\bibnamefont
  {Georges}},\ }\bibfield  {title} {\enquote {\bibinfo {title} {Anomalous
  diffusion in disordered media: statistical mechanisms, models and physical
  applications},}\ }\href@noop {} {\bibfield  {journal} {\bibinfo  {journal}
  {Phys. Rep.}\ }\textbf {\bibinfo {volume} {195}},\ \bibinfo {pages}
  {127--293} (\bibinfo {year} {1990})}\BibitemShut {NoStop}%
\bibitem [{\citenamefont {Metzler}\ and\ \citenamefont
  {Klafter}(2000)}]{Metzler2000PR}%
  \BibitemOpen
  \bibfield  {author} {\bibinfo {author} {\bibfnamefont {R.}~\bibnamefont
  {Metzler}}\ and\ \bibinfo {author} {\bibfnamefont {J.}~\bibnamefont
  {Klafter}},\ }\bibfield  {title} {\enquote {\bibinfo {title} {The random
  walk's guide to anomalous diffusion: a fractional dynamics approach},}\
  }\href@noop {} {\bibfield  {journal} {\bibinfo  {journal} {Phys. Rep.}\
  }\textbf {\bibinfo {volume} {339}},\ \bibinfo {pages} {1--77} (\bibinfo
  {year} {2000})}\BibitemShut {NoStop}%
\bibitem [{\citenamefont {Vlahos}\ and\ \citenamefont
  {Isliker}(2008)}]{Vlahos2008arXiv}%
  \BibitemOpen
  \bibfield  {author} {\bibinfo {author} {\bibfnamefont {L.}~\bibnamefont
  {Vlahos}}\ and\ \bibinfo {author} {\bibfnamefont {H.}~\bibnamefont
  {Isliker}},\ }\bibfield  {title} {\enquote {\bibinfo {title} {Normal and
  anomalous diffusion: A tutorial},}\ }\href@noop {} {\bibfield  {journal}
  {\bibinfo  {journal} {arxiv}\ ,\ \bibinfo {pages} {0805.0419}} (\bibinfo
  {year} {2008})}\BibitemShut {NoStop}%
\bibitem [{\citenamefont {Xia}\ \emph {et~al.}(2014)\citenamefont {Xia},
  \citenamefont {Francois}, \citenamefont {Punzmann},\ and\ \citenamefont
  {Shats}}]{Xia2014PRL}%
  \BibitemOpen
  \bibfield  {author} {\bibinfo {author} {\bibfnamefont {H.}~\bibnamefont
  {Xia}}, \bibinfo {author} {\bibfnamefont {N.}~\bibnamefont {Francois}},
  \bibinfo {author} {\bibfnamefont {H.}~\bibnamefont {Punzmann}}, \ and\
  \bibinfo {author} {\bibfnamefont {M.}~\bibnamefont {Shats}},\ }\bibfield
  {title} {\enquote {\bibinfo {title} {Taylor particle dispersion during
  transition to fully developed two-dimensional turbulence},}\ }\href@noop {}
  {\bibfield  {journal} {\bibinfo  {journal} {Phys. Rev. Lett.}\ }\textbf
  {\bibinfo {volume} {112}},\ \bibinfo {pages} {104501} (\bibinfo {year}
  {2014})}\BibitemShut {NoStop}%
\bibitem [{\citenamefont {Ariel}\ \emph {et~al.}(2015)\citenamefont {Ariel},
  \citenamefont {Rabani}, \citenamefont {Benisty}, \citenamefont {Partridge},
  \citenamefont {Harshey},\ and\ \citenamefont {Be\'er}}]{Ariel2015NC}%
  \BibitemOpen
  \bibfield  {author} {\bibinfo {author} {\bibfnamefont {G.}~\bibnamefont
  {Ariel}}, \bibinfo {author} {\bibfnamefont {A.}~\bibnamefont {Rabani}},
  \bibinfo {author} {\bibfnamefont {S.}~\bibnamefont {Benisty}}, \bibinfo
  {author} {\bibfnamefont {J.~D.}\ \bibnamefont {Partridge}}, \bibinfo {author}
  {\bibfnamefont {R.~M.}\ \bibnamefont {Harshey}}, \ and\ \bibinfo {author}
  {\bibfnamefont {A.}~\bibnamefont {Be\'er}},\ }\bibfield  {title} {\enquote
  {\bibinfo {title} {Swarming bacteria migrate by {L\'evy} walk},}\ }\href@noop
  {} {\bibfield  {journal} {\bibinfo  {journal} {Nat. Comm.}\ }\textbf
  {\bibinfo {volume} {6}},\ \bibinfo {pages} {8396} (\bibinfo {year}
  {2015})}\BibitemShut {NoStop}%
\bibitem [{\citenamefont {Jullien}, \citenamefont {Paret},\ and\ \citenamefont
  {Tabeling}(1999)}]{Jullien1999PRL}%
  \BibitemOpen
  \bibfield  {author} {\bibinfo {author} {\bibfnamefont {M.-C.}\ \bibnamefont
  {Jullien}}, \bibinfo {author} {\bibfnamefont {J.}~\bibnamefont {Paret}}, \
  and\ \bibinfo {author} {\bibfnamefont {P.}~\bibnamefont {Tabeling}},\
  }\bibfield  {title} {\enquote {\bibinfo {title} {Richardson pair dispersion
  in two-dimensional turbulence},}\ }\href@noop {} {\bibfield  {journal}
  {\bibinfo  {journal} {Phys. Rev. Lett.}\ }\textbf {\bibinfo {volume} {82}},\
  \bibinfo {pages} {2872} (\bibinfo {year} {1999})}\BibitemShut {NoStop}%
\bibitem [{\citenamefont {Falkovich}\ \emph {et~al.}(2012)\citenamefont
  {Falkovich}, \citenamefont {Xu}, \citenamefont {Pumir}, \citenamefont
  {Bodenschatz}, \citenamefont {Biferale}, \citenamefont {Boffetta},
  \citenamefont {Lanotte},\ and\ \citenamefont {Toschi}}]{Falkovich2012PoF}%
  \BibitemOpen
  \bibfield  {author} {\bibinfo {author} {\bibfnamefont {G.}~\bibnamefont
  {Falkovich}}, \bibinfo {author} {\bibfnamefont {H.}~\bibnamefont {Xu}},
  \bibinfo {author} {\bibfnamefont {A.}~\bibnamefont {Pumir}}, \bibinfo
  {author} {\bibfnamefont {E.}~\bibnamefont {Bodenschatz}}, \bibinfo {author}
  {\bibfnamefont {L.}~\bibnamefont {Biferale}}, \bibinfo {author}
  {\bibfnamefont {G.}~\bibnamefont {Boffetta}}, \bibinfo {author}
  {\bibfnamefont {A.}~\bibnamefont {Lanotte}}, \ and\ \bibinfo {author}
  {\bibfnamefont {F.}~\bibnamefont {Toschi}},\ }\bibfield  {title} {\enquote
  {\bibinfo {title} {On {Lagrangian} single-particle statistics},}\ }\href@noop
  {} {\bibfield  {journal} {\bibinfo  {journal} {Phys. Fluids}\ }\textbf
  {\bibinfo {volume} {24}},\ \bibinfo {pages} {055102} (\bibinfo {year}
  {2012})}\BibitemShut {NoStop}%
\bibitem [{\citenamefont {Huang}\ \emph {et~al.}(2013)\citenamefont {Huang},
  \citenamefont {Biferale}, \citenamefont {Calzavarini}, \citenamefont {Sun},\
  and\ \citenamefont {Toschi}}]{Huang2013PRE}%
  \BibitemOpen
  \bibfield  {author} {\bibinfo {author} {\bibfnamefont {Y.}~\bibnamefont
  {Huang}}, \bibinfo {author} {\bibfnamefont {L.}~\bibnamefont {Biferale}},
  \bibinfo {author} {\bibfnamefont {E.}~\bibnamefont {Calzavarini}}, \bibinfo
  {author} {\bibfnamefont {C.}~\bibnamefont {Sun}}, \ and\ \bibinfo {author}
  {\bibfnamefont {F.}~\bibnamefont {Toschi}},\ }\bibfield  {title} {\enquote
  {\bibinfo {title} {{Lagrangian single particle turbulent statistics through
  the Hilbert-Huang Transforms}},}\ }\href@noop {} {\bibfield  {journal}
  {\bibinfo  {journal} {Phys. Rev. E}\ }\textbf {\bibinfo {volume} {87}},\
  \bibinfo {pages} {041003(R)} (\bibinfo {year} {2013})}\BibitemShut {NoStop}%
\bibitem [{\citenamefont {Bramwell}, \citenamefont {Holdsworth},\ and\
  \citenamefont {Pinton}(1998)}]{Bramwell1998Nature}%
  \BibitemOpen
  \bibfield  {author} {\bibinfo {author} {\bibfnamefont {S.}~\bibnamefont
  {Bramwell}}, \bibinfo {author} {\bibfnamefont {P.}~\bibnamefont
  {Holdsworth}}, \ and\ \bibinfo {author} {\bibfnamefont {J.-F.}\ \bibnamefont
  {Pinton}},\ }\bibfield  {title} {\enquote {\bibinfo {title} {Universality of
  rare fluctuations in turbulence and critical phenomena},}\ }\href@noop {}
  {\bibfield  {journal} {\bibinfo  {journal} {Nature}\ }\textbf {\bibinfo
  {volume} {396}},\ \bibinfo {pages} {552--554} (\bibinfo {year}
  {1998})}\BibitemShut {NoStop}%
\bibitem [{\citenamefont {Bramwell}\ \emph {et~al.}(2000)\citenamefont
  {Bramwell}, \citenamefont {Christensen}, \citenamefont {Fortin},
  \citenamefont {Holdsworth}, \citenamefont {Jensen}, \citenamefont {Lise},
  \citenamefont {L{\'o}pez}, \citenamefont {Nicodemi}, \citenamefont {Pinton},\
  and\ \citenamefont {Sellitto}}]{Bramwell2000PRL}%
  \BibitemOpen
  \bibfield  {author} {\bibinfo {author} {\bibfnamefont {S.}~\bibnamefont
  {Bramwell}}, \bibinfo {author} {\bibfnamefont {K.}~\bibnamefont
  {Christensen}}, \bibinfo {author} {\bibfnamefont {J.-Y.}\ \bibnamefont
  {Fortin}}, \bibinfo {author} {\bibfnamefont {P.}~\bibnamefont {Holdsworth}},
  \bibinfo {author} {\bibfnamefont {H.}~\bibnamefont {Jensen}}, \bibinfo
  {author} {\bibfnamefont {S.}~\bibnamefont {Lise}}, \bibinfo {author}
  {\bibfnamefont {J.}~\bibnamefont {L{\'o}pez}}, \bibinfo {author}
  {\bibfnamefont {M.}~\bibnamefont {Nicodemi}}, \bibinfo {author}
  {\bibfnamefont {J.-F.}\ \bibnamefont {Pinton}}, \ and\ \bibinfo {author}
  {\bibfnamefont {M.}~\bibnamefont {Sellitto}},\ }\bibfield  {title} {\enquote
  {\bibinfo {title} {Universal fluctuations in correlated systems},}\
  }\href@noop {} {\bibfield  {journal} {\bibinfo  {journal} {Phys. Rev. Lett.}\
  }\textbf {\bibinfo {volume} {84}},\ \bibinfo {pages} {3744} (\bibinfo {year}
  {2000})}\BibitemShut {NoStop}%
\bibitem [{\citenamefont {Li}\ and\ \citenamefont
  {Huang}(2014)}]{Li2014PhysicaA}%
  \BibitemOpen
  \bibfield  {author} {\bibinfo {author} {\bibfnamefont {M.}~\bibnamefont
  {Li}}\ and\ \bibinfo {author} {\bibfnamefont {Y.}~\bibnamefont {Huang}},\
  }\bibfield  {title} {\enquote {\bibinfo {title} {{Hilbert--Huang Transform
  based multifractal analysis of China stock market}},}\ }\href@noop {}
  {\bibfield  {journal} {\bibinfo  {journal} {Physica A}\ }\textbf {\bibinfo
  {volume} {406}},\ \bibinfo {pages} {222--229} (\bibinfo {year}
  {2014})}\BibitemShut {NoStop}%
\bibitem [{Note2()}]{Note2}%
  \BibitemOpen
  \bibinfo {note} {See
  {http://damtp.cam.ac.uk/user/gold/datarequests.html}}\BibitemShut {NoStop}%
\bibitem [{Note3()}]{Note3}%
  \BibitemOpen
  \bibinfo {note} {See {https://github.com/lanlankai}}\BibitemShut {NoStop}%
\end{thebibliography}
%

\end{document}